\documentclass[10pt,conference]{IEEEtran}

\usepackage{booktabs}   %% For formal tables:
\usepackage{subcaption} %% For complex figures with subfigures/subcaptions
\usepackage[utf8]{inputenc} % allow utf-8 input
\usepackage[T1]{fontenc}    % use 8-bit T1 fonts
\usepackage[hidelinks]{hyperref}     % hyperlinks
\usepackage{url}            % simple URL typesetting
\usepackage{booktabs}       % professional-quality tables
\usepackage{amsfonts}       % blackboard math symbols
\usepackage{microtype}      % microtypography
\usepackage{xcolor}
\usepackage{float} 
\usepackage{fancyhdr}

\usepackage{graphicx}
\usepackage{multirow}
\usepackage{cite}
\usepackage{amsmath}

\usepackage{listings}

\usepackage{algorithm}
\usepackage{algorithmicx}
\usepackage{algpseudocode}
\MakeRobust{\Call}
\floatname{algorithm}{Algorithm}

\algnewcommand{\LeftComment}[1]{\Statex \(\triangleright\) #1}

\lstdefinestyle{mystyle}{
    frame=tb,
    commentstyle=\color{codegreen},
    keywordstyle=\color{magenta},
    numberstyle=\tiny\color{codegray},
    stringstyle=\color{codepurple},
    basicstyle={\scriptsize\ttfamily},
    breakatwhitespace=false,         
    breaklines=true,                 
    captionpos=b,                    
    keepspaces=true,                 
    numbers=left,                    
    numbersep=5pt,                  
    showspaces=false,                
    showstringspaces=false,
    showtabs=false,                  
    tabsize=2,
    xleftmargin=1ex,
    belowskip=-0.5em,
    aboveskip=0.5em,
}
\usepackage{amssymb}% http://ctan.org/pkg/amssymb
\usepackage{pifont}% http://ctan.org/pkg/pifont
\newcommand{\cmark}{\ding{51}}%
\newcommand{\xmark}{\ding{55}}%

\definecolor{codegreen}{rgb}{0,0.6,0}
\definecolor{codegray}{rgb}{0.5,0.5,0.5}
\definecolor{codepurple}{rgb}{0.58,0,0.82}
\definecolor{backcolour}{rgb}{0.95,0.95,0.92}
\definecolor{codered}{rgb}{1.0,0.01,0.24}
\definecolor{codeblue}{rgb}{0.0, 0.18, 0.65}

\usepackage{xspace}
\newcommand{\name}{CuPBoP\xspace}

\newcommand{\ignore}[1]{}

\newcommand{\greenv}{\textcolor{codegreen}{\cmark}}
\newcommand{\redx}{\textcolor{codered}{\xmark}}

\usepackage{array}
\newcommand{\PreserveBackslash}[1]{\let\temp=\\#1\let\\=\temp}
\newcolumntype{C}[1]{>{\PreserveBackslash\centering}p{#1}}
\newcolumntype{R}[1]{>{\PreserveBackslash\raggedleft}p{#1}}
\newcolumntype{L}[1]{>{\PreserveBackslash\raggedright}p{#1}}

\begin{document}

\title{\name{}: Cuda for Parallelized and Broad-range Processors}

%\ignore{
\author{
\IEEEauthorblockN{Ruobing Han}
\IEEEauthorblockA{\textit{Georgia Institute of Technology} \\ \textit{hanruobing@gatech.edu}}
\and
\IEEEauthorblockN{Jun Chen$^*$}
\IEEEauthorblockA{\textit{Georgia Institute of Technology} \\ \textit{jchen706@gatech.edu}}
\and
\IEEEauthorblockN{Bhanu Garg$^*$}
\IEEEauthorblockA{\textit{Georgia Institute of Technology} \\ \textit{bgarg@gatech.edu}}
\and
\IEEEauthorblockN{Jeffrey Young}
\IEEEauthorblockA{\textit{Georgia Institute of Technology} \\ \textit{jyoung9@gatech.edu}}
\and
\IEEEauthorblockN{Jaewoong Sim}
\IEEEauthorblockA{\textit{Seoul National University} \\ \textit{jaewoong@snu.ac.kr}}
\and
\IEEEauthorblockN{Hyesoon Kim}
\IEEEauthorblockA{\textit{Georgia Institute of Technology} \\ \textit{hyesoon@cc.gatech.edu}}
}
%}

\IEEEtitleabstractindextext{%
\begin{abstract}
CUDA is one of the most popular choices for GPU programming, but it can only be executed on NVIDIA GPUs. Executing CUDA on non-NVIDIA devices not only benefits the hardware community, but also allows data-parallel computation in heterogeneous systems. To make CUDA programs portable, some researchers have proposed using source-to-source translators to translate CUDA to portable programming languages that can be executed on non-NVIDIA devices. However, most CUDA translators require additional manual modifications on the translated code, which imposes a heavy workload on developers.  \\
%In addition, the rapid development of new CUDA features and libraries mean that these translators consistently struggle to support the latest features in a timely fashion. \\
In this paper, \name{} is proposed to execute CUDA on non-NVIDIA devices without relying on any portable programming languages. Compared with existing work that executes CUDA on non-NVIDIA devices, \name{} does not require manual modification of the CUDA source code, but it still achieves the highest coverage (69.6\%), much higher than existing frameworks (56.6\%) on the Rodinia benchmark. In particular, for CPU backends, \name{} supports several ISAs (e.g., X86, RISC-V, AArch64) and has close or even higher performance compared with other projects. We also compare and analyze the performance among \name{}, manually optimized OpenMP/MPI programs, and CUDA programs on the latest Ampere architecture GPU, and show future directions for supporting CUDA programs on non-NVIDIA devices with high performance.
\end{abstract}

% % Note that keywords are not normally used for peerreview papers.
 
 }

\maketitle
\def\thefootnote{*}\footnotetext{These authors contributed equally to this work}

\IEEEdisplaynontitleabstractindextext

\IEEEpeerreviewmaketitle

\section{Introduction}
As high-performance computing (HPC) platforms have diversified, the possibilities for execution have grown from conventional CPUs to highly vectorized CPUs, GPUs developed by NVIDIA, AMD, and Intel and even newer high-performance CPUs from Arm and emerging RISC-V platforms.  
Although OpenCL, SYCL, and portability APIs like Kokkos~\cite{kokkos,CarterEdwards20143202} and RAJA~\cite{raja} have all been proposed as methods to support this new era of extreme heterogeneous HPC hardware, none of these frameworks or APIs have proved to be quite as performant as CUDA codes written for NVIDIA GPUs~\cite{perkins2017cuda}. This means that CUDA is still currently the de-facto choice for GPU programming for HPC applications. CUDA is a proprietary language that cannot be executed on non-NVIDIA devices. If we can eliminate this limitation, then non-NVIDIA devices can benefit from the rapid development of CUDA applications. It can also benefit single-kernel multiple-device (SKMD) systems that rely on data-parallel computation~\cite{lee2013transparent,7855896,pandit2014fluidic} as a part of heterogeneous systems.

To leverage the large code base of CUDA HPC applications on non-NVIDIA devices, some developers have tried to manually rewrite CUDA programs using portable languages~\cite{gu2016opencl,tschopp2016efficient, perkins2016cltorch,daley2020case,bercea2015performance}. 
However, manual porting is not a viable approach in the long run because of the rapid application development in the CUDA community. To reduce the effort spent on code porting, researchers~\cite{perkins2017cuda, harvey2011swan} have proposed using source-to-source translators to automatically translate CUDA to portable languages. However, CUDA is based on C++, which may contain complex macro definitions. While research compilers like ROSE~\cite{quinlan:2019:rose} can potentially handle these macros, most existing industry-supported
translators like AMD's HIPIFY~\cite{HIPIFY} and Intel's DPCT~\cite{DPCT} cannot correctly translate them~\cite{castano2021intel}. Thus, translated CUDA codes may require some manual modifications by HPC software developers for each non-CUDA platform that is targeted.

Based on these observations, we envision that it is necessary to directly parse and execute the CUDA source code at an appropriate intermediate representation level instead of relying on portable languages. In this paper, a new framework, \name{}, is proposed to support executing CUDA source code on non-NVIDIA architectures. CuPBop is the first approach that allows for fully compiler-automated SPMD to MPMD code conversion, which enables running CUDA codes on a wide variety of CPU-based platforms and ISAs (x86, aarch64, RISC-V) with high performance and code coverage.
%The workflow for \name{} is shown in Figure ~\ref{fig:cupbop_pipeline}. \name{} consists of two parts: compilation and runtime. The compilation part compiles the input CUDA programs to LLVM IR (bytecode), applies the transformations accordingly, and generates binary object files for a given ISA. The runtime part consists of the libraries that implement several CUDA built-in functions used to control the interaction between the host and the devices, such as kernel launch, memory allocation and memory movement.

%Specifically, in this paper, we focus how to efficiently support CPU backends with \name{}. Insights developed from supporting CUDA's SPMD on other architectures such as non-NVIDIA GPUs and CPUs can also be used to effectively map these codes to future Processing Near Memory like accelerators~\cite{samsung-pim} or FPGAs. 

Compared with existing solutions  ~\cite{HIP-CPU,reinders2020c++,han2021cox,blomkvist2021cumulus}, \name{} doesn't require any manual modifications to CUDA source code, and it also achieves the highest coverage on HPC and database benchmarks. For the Rodinia benchmark suite, \name{}'s coverage (69.8\%) is much higher than existing frameworks (56.6\%). \par
\ignore{
Compared with existing solutions~\cite{HIP-CPU,reinders2020c++,han2021cox,blomkvist2021cumulus}, \name{} has the following unique features: \textcolor{blue}{ 1) \name allows for executing CUDA source code without programmer intervention, even with complex macro usage;} 2) Codes compiled with \name{} achieve close or better performance compared with existing solutions; and 3) \name{}'s use of LLVM and NVVM IRs allows for better support for different LLVM backends, including AArch64 and RISC-V ISAs. New contributions of this work include: 
}
In this paper, the following technical contributions are made:
\begin{itemize}
    %\item Analyzes the existing solutions for executing CUDA on CPU architectures;
    \item A detailed exploration of \name{} compilation and runtime components which efficiently execute CUDA codes on multi-core CPU platforms without programmer modification. 
     \item Comparison of code performance and coverage with two state-of-the art source-to-source translator tools, AMD's HIPIFY and Intel's DPCT;
    \item Analysis of the performance for running CUDA programs from Rodinia and Hetero-Mark on five different CPU platforms, including Intel Cascade Lake, AMD Milan, Arm A64FX, Arm Ampere, and SiFive RISC-V CPUs; 
    \item Discussion of the roofline model for several of these CPU platforms compared to a baseline NVIDIA Ampere GPU as well as needed optimizations to close the performance gap between transformed CPU codes and CUDA codes.
%    \item Proposes optimizations that are useful for improving the performance on executing CUDA programs on CPU;
%    \item Evaluate \name{} with 2 source-to-source translator used in industry;
%    \item Analyze the performance gap between manually optimized CPU code and code translated from CUDA.
    %\item Evaluates \name{} on 2 benchmarks (Hetero-Mark, Rodinia) on 3 CPU processors (x86, AArch64, RISC-V);
    %\item Compare \name{} performance with the latest NVIDIA Ampere GPU, and use roofline~\cite{williams2009roofline} model to analyze the results;
    %\item Compare the performance between \name{} and NVIDIA-GPU;
\end{itemize}

\ignore{
  \begin{figure}[htbp]
    \centering
    \includegraphics[width=70mm]{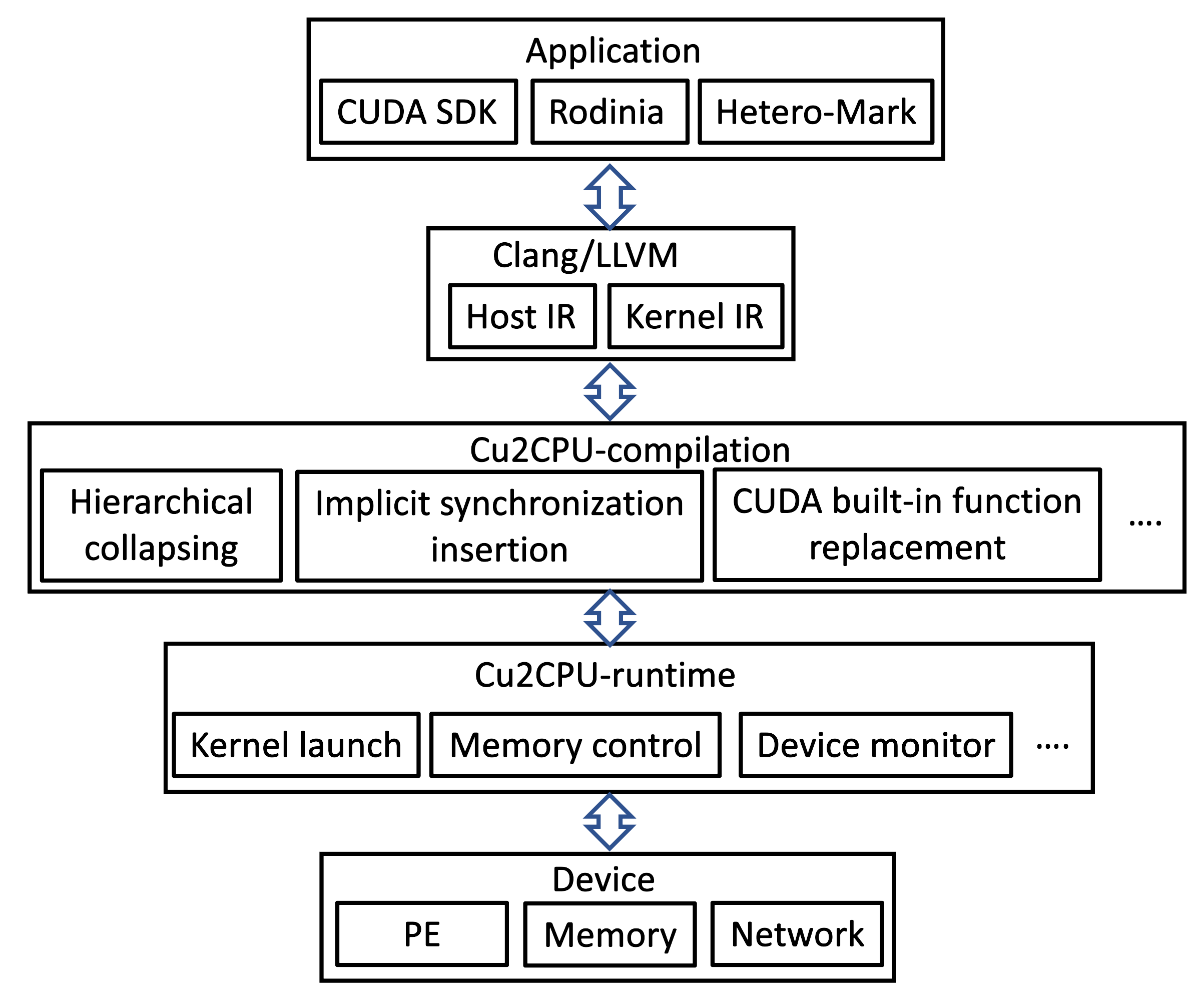}
    \caption{The hierarchical abstract for \name{}.}
    \label{fig:cupbop_pipeline_arch}
  \end{figure} 
 }

\section{Background~\label{sec:background}}

\begin{figure*}[h!]
    \centering
    \includegraphics[width=150mm]{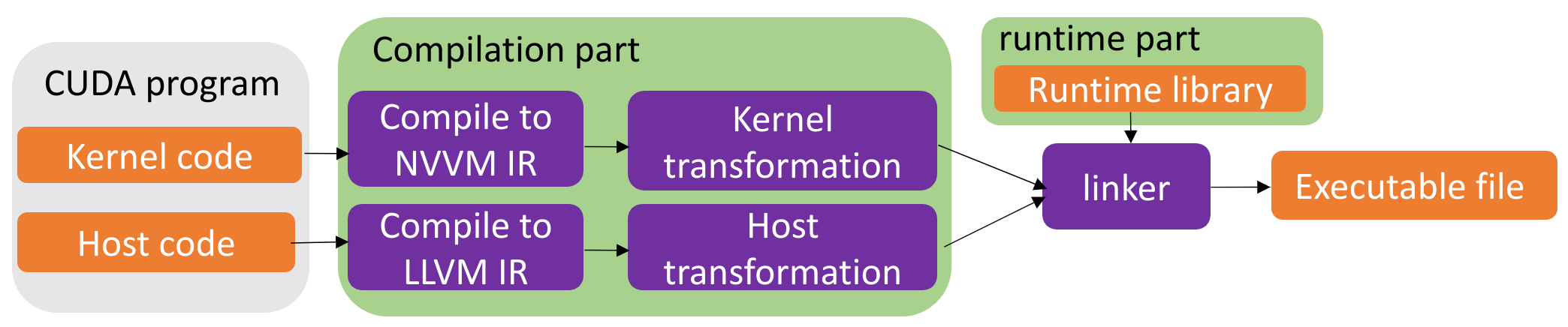}
    \caption{The workflow for generating non-NVIDIA device executable files from CUDA source code.}
    \label{fig:cupbop_pipeline}
  \end{figure*}

The workflow from CUDA source code to executable files for a non-NVIDIA backend is shown in Figure~\ref{fig:cupbop_pipeline}. The entire \name workflow consists of a compilation and runtime phase. The compilation phase uses two separate paths to compile the CUDA kernel/host programs to binary files for a given backend, and the runtime phase contains libraries that can be linked with generated binaries to implement CUDA runtime functions for non-NVIDIA devices.
%During the process, CUDA kernels, which are SPMD formats, are translated to MPMD formats. In addition, CUDA has different memory space, like global memory, shared memory and constant memory. As there are no exact counterparts in CPU, memory space modifications are needed. The discussion of all these challenges and solutions are in Section.~\ref{sec:compilation}.
%These libraries can be linked with the binary files that are generated by the compilation part to produce the executable files for the given backend. 

\begin{figure}[htbp]
    \centering
    \includegraphics[width=90mm]{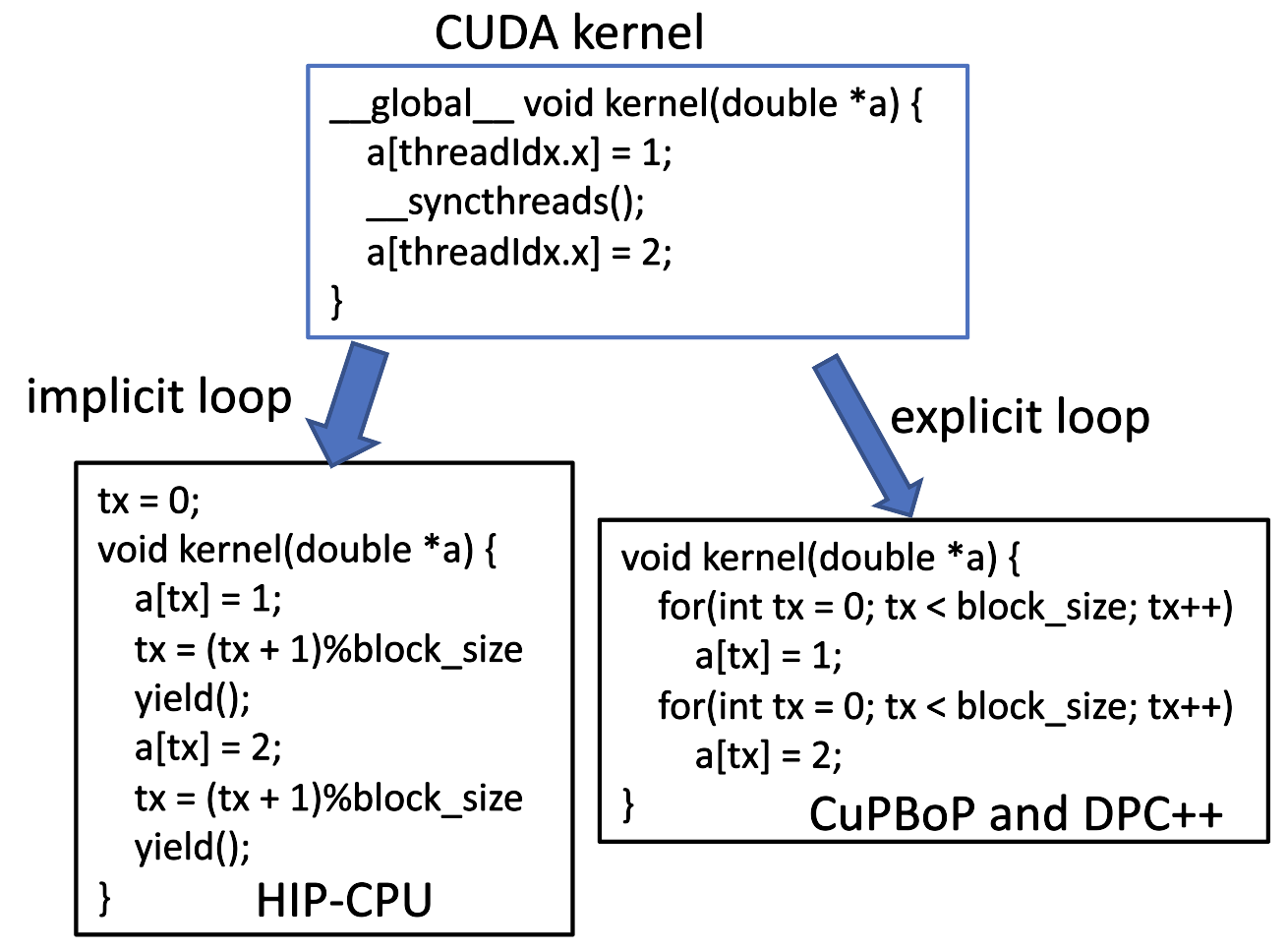}
    \caption{Support SPMD-to-MPMD transformation with implicit/explicit loops.}
    \label{fig:support_sync}
  \end{figure}

The main compilation tasks supported by \name focus on supporting varying thread-level parallelism, memory hierarchies like shared and global memory, and synchronization (e.g., {\tt sync\_threads()}, implicit synchronizations across kernels) across heterogeneous architectures. CPUs and current-generation Intel GPUs can execute many fewer threads than NVIDIA GPUs, so researchers have proposed using block-fusion~\cite{stratton2008mcuda, stratton2010efficient} to wrap a GPU block's workload into a single function and execute it on a CPU core. As an example, a GPU vector addition kernel in Listing~\ref{code:gpu_program} that has $gridSize * blockSize$ number of threads is converted into  Listing~\ref{code:cpu_program} that only invokes the kernel {\tt gridSize} times, possibly in a parallel fashion. We refer to this transformation as Single Program Multiple Data to Multiple Program Multiple Data (SPMD-to-MPMD) in this paper. To the best of our knowledge, two mechanisms support SPMD-to-MPMD transformation. An example is shown in Figure~\ref{fig:support_sync}. Both mechanisms use loops to warp the original SPMD functions. HIP-CPU uses fiber to implement implicit loops, while \name{} and DPC++ use explicit for-loops. To support CUDA {\_\_syncthread}, HIP-CPU uses yield instructions which switch to the next fiber, while \name{} and DPC++ use loop fission.

%(2) Supporting different memory hierarchy e.g.) shared memory, global memory, and (3)supporting different level of synchronizations (e.g., {\tt synch\_threads()}, implicit synchronizations across kernels. 
%

%The kernel transformation and host transformation are described in Section~\ref{sec:compilation}, while the runtime library is discussed in Section~\ref{sec:runtime}.

%Although CUDA is based on C/C++ that can be executed on CPU with bearable manually modifications, the challenge to execute CUDA programs on CPU is the parallelism gap between GPUs and CPUs. The modern GPUs support executing thousands of GPU threads. As for a single CPU, it can hardly support more than one hundred threads simultaneously. To elimination the gap, some researchers suggest to use block-fusion~\cite{stratton2008mcuda, stratton2010efficient} to wrap a GPU block's workload into a single function and execute it by a CPU core. In other words, each GPU Streaming multiprocessors (SM) is mapped to a CPU core. For example, the original GPU program for Vector Add and the translated CPU program are shown in Code.~\ref{code:gpu_program} and Code.~\ref{code:cpu_program}. Before the translation, there are $gridSize * blockSize$ GPU threads. After translation, CPU only needs to invoke the wrapped kernel {\tt gridSize} times. This transformation is called SPMD-to-MPMD transformation in this paper

.%Although this transformation looks simple, some cases are needed to concern. For GPU kernel, there are many features needed to support, like shared memory, constant memory, warp shuffle. The implementation of these GPU features on CPU is discussed in Section~\ref{sec:compilation}. To utilize multi-core CPU, the for-loop in Code.~\ref{code:cpu_program} should be replaced. Besides, there are some CUDA runtime features (e.g., kernel launch, device synchronization) also require special concern to be supported on CPU. The detail is recorded in Section~\ref{sec:runtime}.

\begin{lstlisting}[caption={Original CUDA GPU program.},label={code:gpu_program},language=C]
__global__ void vecAdd(double *a, double *b, double *c, int n) {
    int id = blockIdx.x*blockDim.x+threadIdx.x;
    c[id] = a[id] + b[id];
}

int main() {
    ...
    int gridSize = 2; int blockSize = 2;
    vecadd<<<gridSize, blockSize>>>(d_a, d_b, d_c, n);
}
\end{lstlisting}

\begin{lstlisting}[caption={Translated CPU program.},label={code:cpu_program},language=C]
int blockIdx, blockDim;
void vecAdd(double *a, double *b, double *c, int n) {
    for(int tid = 0; tid<blockDim; tid++) {
        int id = blockIdx*blockDim+tid;
        c[id] = a[id] + b[id];
    }
}
int main() {
    ...
    int gridSize = 2; int blockSize = 2;
    // need to be replaced by multi-thread
    for(int bid = 0; bid<gridSize; bid++) {
        blockIdx = bid; blockDim = gridSize;
        vecadd(d_a, d_b, d_c, n);
    }
}
\end{lstlisting}

%Besides the above transformation, to execute CUDA on non-NVIDIA devices, the developers are required to implement CUDA runtime libraries on non-NVIDIA devices. 
\name{} implements the SPMD-to-MPMD transformation proposed in COX~\cite{han2021cox}. In COX, the author propose a new SPMD-to-MPMD algorithm that supports warp-level functions that are newly introduced in CUDA 9.1. \par
However, to execute unmodified CUDA programs, the CUDA host code also needs to be supported. In Listing~\ref{code:cpu_program} lines 13-16, we show a naive example by using a CPU function call to support the CUDA kernel launch. To fully utilize the CPU computation resource, 
efficient concurrent processing is required. These runtime system contents, which are orthogonal with SPMD-to-MPMD transformation, are not  covered in COX.

\begin{figure}[htbp]
    \centering
    \includegraphics[width=90mm]{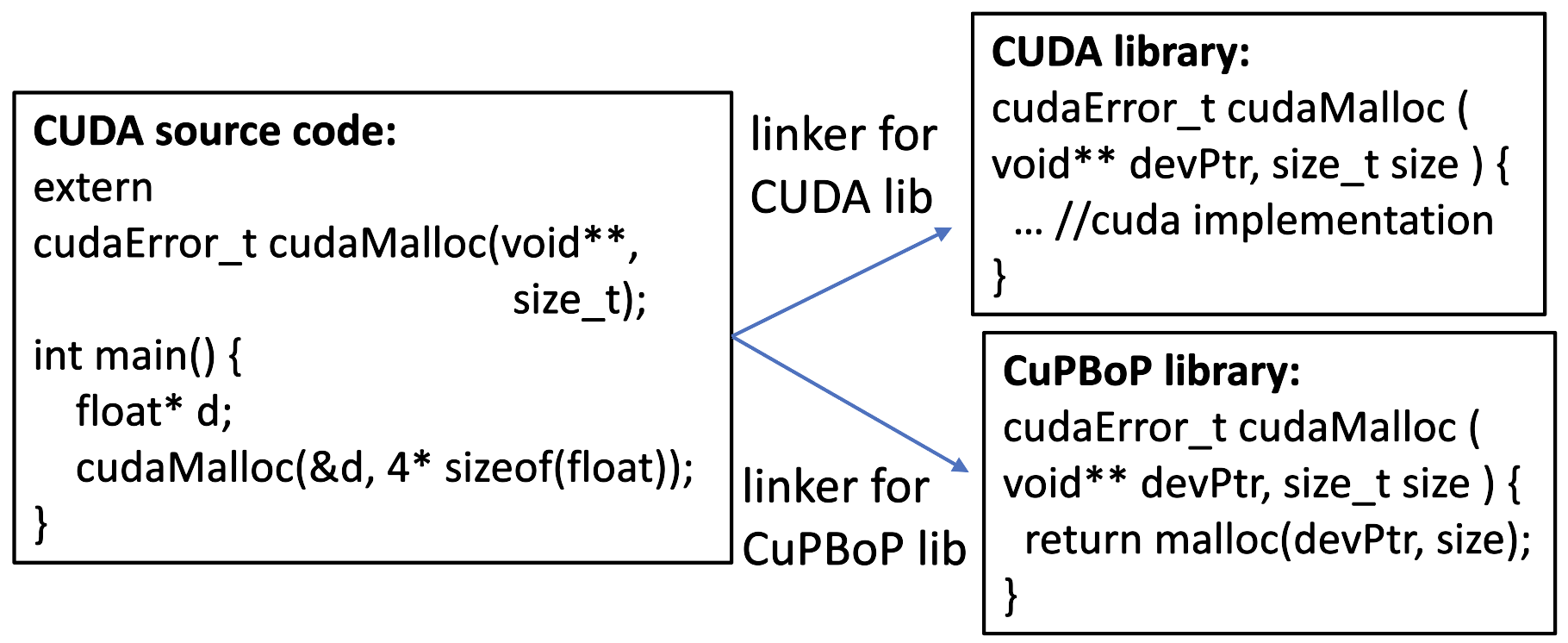}
    \caption{By linking to different libraries, a program can be executed on both NVIDIA GPUs and CPUs without modification.}
    \label{fig:runtime_explain}
  \end{figure}

The main tasks of the runtime system are to support data movement among host and device memory regions, kernel launches, and related functionality. Figure~\ref{fig:runtime_explain} shows an example of runtime library replacement for a device-specific version of the {\tt cudaMalloc} function. When the host and CUDA kernels are running in the CPU memory space, {\tt cudaMalloc} is simply replaced by {\tt malloc} using a \name library. However, if the host invokes a kernel for NVIDIA GPU, the traditional CUDA runtime library is called transparently. By changing the libraries to be linked, a program can be executed on both NVIDIA GPUs and CPUs without modifications to the input code or to the libraries used to compile the binary. 

\section{Compilation\label{sec:compilation}}
In this section, a simple CUDA example (Listing~\ref{code:cuda_reverse}) is used to explain \name's approach for compiler-level transformation. 
\begin{lstlisting}[caption={CUDA program, with dynamic shared memory.},label={code:cuda_reverse},language=C]
__global__ void dynamicReverse(int *d, int n) {
  extern __shared__ int s[];
  int t = threadIdx.x;
  int tr = n - t - 1;
  s[t] = d[t];
  __syncthreads();
  d[t] = s[tr];
}

int main() {
  ...
  cudaMalloc(&d_d, n * sizeof(int));
  dynamicReverse<<<1, n, n * sizeof(int)>>>(d_d, n);
  cudaMemcpy(d, d_d, n * sizeof(int), cudaMemcpyDeviceToHost);
  ...
}
\end{lstlisting}

%demonstrate how \name{} applies compiler-level transformations to support executing CUDA on CPU devices. \\
The CUDA program consists of two functions: the host function ({\tt main}) and the kernel function ({\tt dynamicReverse}). The kernel function declares an external variable ({\tt s}) that is located in GPU shared memory. The size of this variable ({\tt n * sizeof(int)}) is defined at runtime. There is a block-level barrier ({\tt \_\_syncthreads()}) inside the kernel function. 
%Within each GPU block, threads can not execute instructions after the barrier until all threads have reached the barrier. \\
%The host function uses CUDA runtime functions (e.g., {\tt cudaMalloc, cudaMemcpy}) to manage the GPU memory and launch the kernel.

\ignore{
  \begin{figure}[htbp]
    \centering
    \includegraphics[width=80mm]{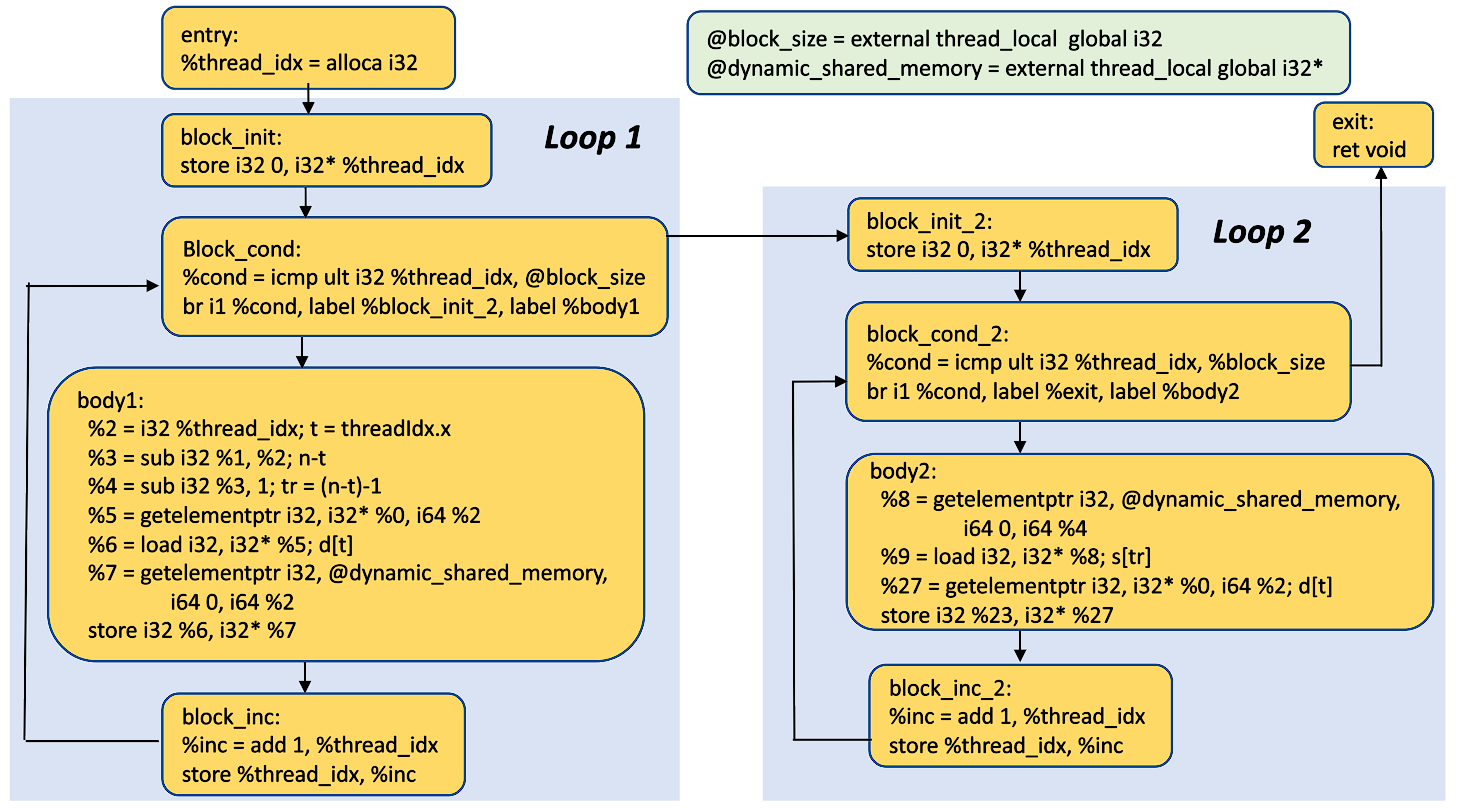}
    \caption{\name Transformed LLVM IR from Listing~\ref{code:cuda_reverse}.}
    \label{fig:generated_kernel}
  \end{figure} 
  
The LLVM IR generated by \name{} is shown in Figure~\ref{fig:generated_kernel}. The following subsection describe steps for transforming Listing~\ref{code:cuda_reverse} to LLVM IR (Figure~\ref{fig:generated_kernel}).
}

\subsection{Compilation to NVVM \& LLVM IR}
As shown in Figure~\ref{fig:cupbop_pipeline}, \name{} uses Clang to compile the CUDA source code into IRs. The host code and kernel code are compiled separately into two different intermediate representations (IRs): NVVM IR for the CUDA kernel code and LLVM IR for the host code. This additional pass to generate NVVM IR allows for additional compilation transformation capabilities of CUDA inputs that cannot easily be managed with header-based tools like HIP-CPU.
%The whole compilation process can be done without NVCC or CUDA toolkit.
\ignore{
The generated NVVM IR is shown in Listing~\ref{code:nvvm_ir}\footnote{This IR has been pruned for simplicity.};
%declare void @llvm.nvvm.barrier0()
%declare i32 %@llvm.nvvm.read.ptx.sreg.tid.x()
}

\subsection{Transformation of kernel programs (NVVM IR)\label{sec:transform_nvvm_ir}}
After generating the NVVM IR, several transformations are applied to create LLVM IR in an MPMD format (Figure~\ref{fig:generated_kernel}). The following subsections describe these transformations. 
  \begin{figure}[htbp]
    \centering
    \includegraphics[width=80mm]{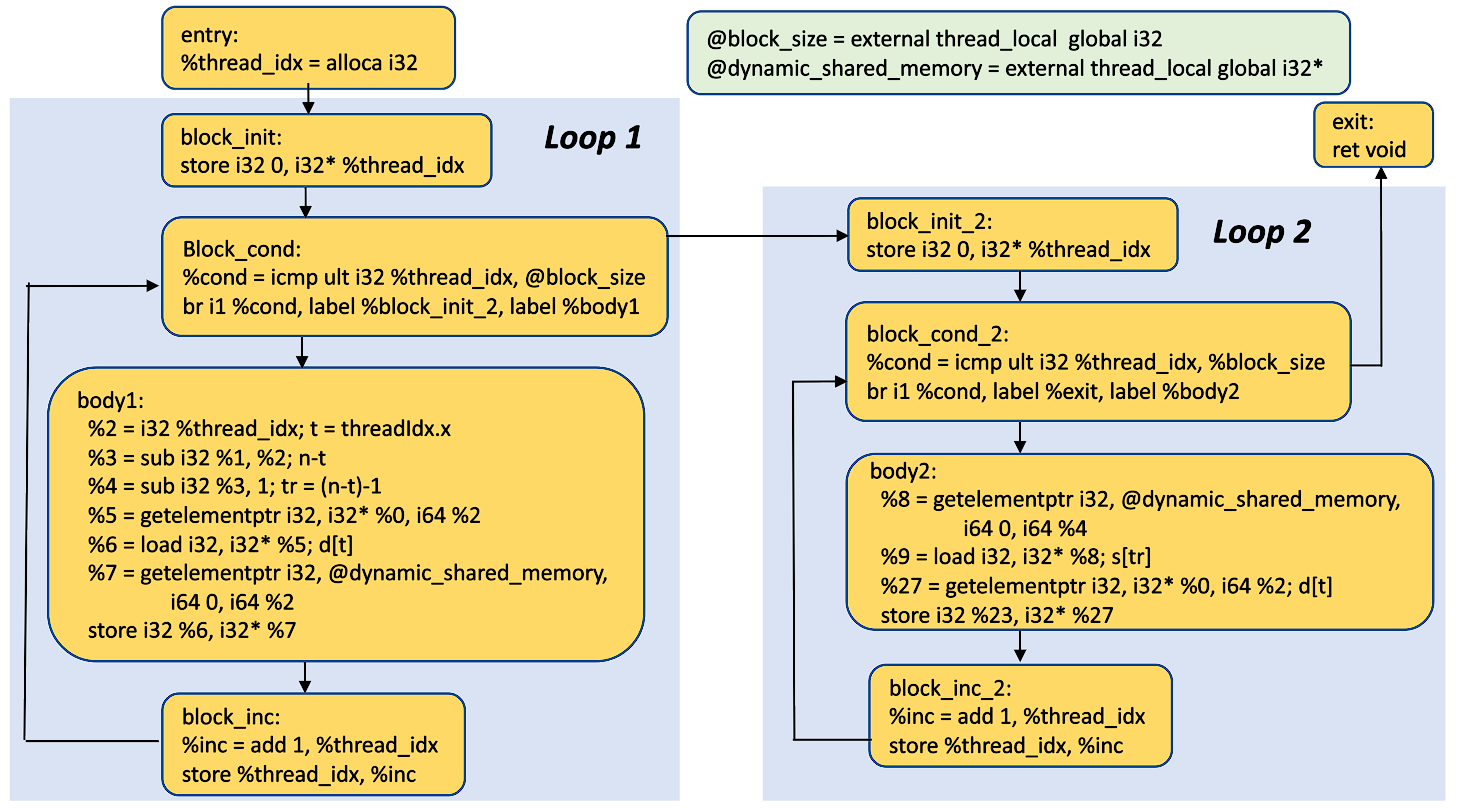}
    \caption{\name transformed LLVM IR from Listing~\ref{code:cuda_reverse}.}
    \label{fig:generated_kernel}
  \end{figure}

\subsubsection{Memory mapping}
CUDA has several types of memory.
%,, however, for CPU devices, only a single memory type is supported. Thus, \name{} will map all CUDA memory to the same CPU memory. However, some memory is shared among all blocks (e.g. global memory, constant memory), 
Some memory addresses are private for single threads (e.g., register), while others are shared within a block (e.g., shared memory) or between blocks (e.g., global memory). \name{} handles these varied memory regions by mapping CUDA global memory to the CPU's heap memory and other CUDA memory spaces to the CPU's stack memory. 
\ignore{
The map between CUDA memory space and CPU memory space is shown in Table~\ref{table:memory_mapping}.
\begin{table}[htbp]
\centering
\begin{tabular}{|c|c|}
\hline
CUDA memory     & CPU memory \\ \hline
global memory   & heap       \\ \hline
share memory    & stack      \\ \hline
private memory  & stack      \\ \hline
constant memory & heap       \\ \hline
texture memory  & heap (WIP) \\ \hline
\end{tabular}
\caption{\name{} uses different CPU memory to support CUDA's hierarchical memory.}
\label{table:memory_mapping}
\end{table}
}
For the given example, the shared memory variable {\tt s} is replaced by a stack memory variable (thread local variable) {\tt dynamic\_shared\_memory} (the green box in Figure~\ref{fig:generated_kernel}).

\subsubsection{Extra variable insertion}
NVIDIA GPUs have a number of special registers~\cite{NVregisterdoc} to support some CUDA intrinsic functions. For example, it uses the register {\tt ctaid} to store the block dimension, which does not have an equivalent in CPU architectures. For this reason, \name{} explicitly declares these variables in the kernel code, and the assignment of the variables is done at runtime. For the given example, a new global variable {\tt block\_size} is created, which is mapped to the CPU stack memory, as GPU blocks from different kernels may have unique block sizes.
\ignore{
\subsection{Metadata replacement}
Metadata is used to generate hardware specific binary files. The input NNVM IR has its metadata for NVIDIA GPU. To generate binary files for CPU, \name{} replaces the metadata.
}

\subsubsection{SPMD to MPMD~\cite{stratton2008mcuda,han2021cox}}
In this step, for-loops are used to wrap the input kernel functions, so that we can compress the whole workloads within a CUDA block into a single function. The transformed LLVM IR from Listing~\ref{code:cuda_reverse} is shown in Figure~\ref{fig:generated_kernel}. Because there is a barrier ({\tt \_\_syncthreads}) in the original program, two for-loops ({\tt Loop1, Loop2}) are generated to wrap codes before/after the barrier. For CUDA programs that do not involve CUDA warp-level features (e.g., warp shuffle, warp vote), we use single layer loops, and the loops' length equal to the {\tt block\_size}, as suggested in ~\cite{stratton2008mcuda}. As for CUDA programs that involve warp-level features, as suggested in ~\cite{han2021cox}, we use nested loops to warp each code section. The inner loops' length equal to the warp size (32 for NVIDIA GPUs), while the outer loops' length equal to $\lceil\frac{block\_size}{warp\_size}\rceil$.

\subsection{Transformation of host programs (LLVM IR)}
CUDA host programs are based on C++, so most instructions in CUDA host code can be executed on CPUs directly. The major challenge for supporting CUDA host programs is to support CUDA runtime functions (e.g., CUDA kernel launch) and any CUDA-specific libraries. %In this section, we discuss the transformations applied for the CUDA host programs.
%For CUDA programs, the host function also be executed on CPU devices. And the host program is C/C++ programs which including some CUDA built-in function. Thus, compared with kernel programs, fewer transformations are needed.
\subsubsection{Implicit barrier insertion}
CUDA has a separate CPU host and GPU device. The CPU host executes the CUDA programs and submits tasks to the GPU device. These tasks include kernel launch and memory management ({\tt cudaMalloc, cudaMemcpy}). For the CPU backend where the host and device are the same device, the host thread doesn't need to submit these memory management tasks to devices but can directly execute them by itself. However, this optimization will incur race conditions that do not exist in the original CUDA programs.
As with CUDA, the kernel launch operation in \name{} is asynchronous, and the host thread is not blocked after a kernel is launched. For a pair of consecutive instructions (Listing~\ref{code:two_kernel}), if the first instruction is a kernel launch that writes to variable {\tt d\_c}, and the second instruction is a memory copy that reads the same variable, the race condition may generate incorrect results. \name{} analyzes the host programs and inserts barriers to avoid potential race conditions. %However, due to alias and other challenges, exactly analysis is difficult and out of the scope of this paper.
\begin{lstlisting}[caption={A barrier is needed between kernel launch and the following memory copy.},label={code:two_kernel},language=C]
    // kernel writes d_c
    vecadd<<<gridSize, blockSize>>>(d_a, d_b, d_c, n);
    // a barrier is needed here
    cudaMemcpy(h_c,d_c,bytes, cudaMemcpyDeviceToHost);
\end{lstlisting}

\subsubsection{Parameter packing}
When the host thread executes a kernel launch function, it submits a task to the task queue. As CUDA programs may invoke kernel functions with various numbers of parameters and types, to implement a universal interface, \name{} packs all kernel parameters into a single variable.
%CUDA programs invokes kernel functions with various number of parameters and types. To avoid requiring additional runtime support for variadic functions, 
\par
\name{} applies transformations to insert a prologue before the kernel launch that packs all the parameters into a single pointer-based object that is passed to the kernel. \name{} then inserts another prologue at the start of the kernel functions to unpack the parameter object. \par
%To support all these cases, \name{} inserts prologues in both GPU kernels and kernel launch function. 
These prologues are implemented at the IR level; however, for simplicity, we show the prologues using a C++ source language example. The prologues for Listing~\ref{code:cuda_reverse} are shown in Listing~\ref{code:prologues} (kernel prologue: line 3-4; host prologue: line 12-18).
As the parameters will be shared between the host thread and the thread pool, all parameters should be in heap memory.
\begin{lstlisting}[caption={Inserted prologues for packing/unpacking parameters.},label={code:prologues},language=C]
__global__ void dynamicReverse(int** p) {
  // prologue for unpacking in a kernel side
  int* d = *((int**)p[0]);
  int n = *((int*)p[1]);
  // end of prologue
  // unpack int* d and int n from int** p
  ... // original code
}

int main(void) {
  // prologue for packing
  int** p = malloc(2 * sizeof(int*));
  int **p0 = new int *;
  *p0 = d_d;
  p[0] = (int *)(p0);
  int *p1 = new int;
  *p1 = n;
  ret[1] = (int *)p1;
  // end of prologue, int* d, int n becomes int**p
  dynamicReverse<<<1, n, n * sizeof(int)>>>(p);
}
\end{lstlisting}

%\name{} first analyzes the kernel functions, and get the function signatures (parameters and their types), and inserts prologues accordingly. These signatures is stored and be used to generate prologues for kernel launch in host programs.

\section{Runtime\label{sec:runtime}}
\name's runtime consists of the CPU implementation of the CUDA runtime functions. It also assigns the variables (e.g., block\_size and dynamic\_shared\_memory in Figure~\ref{fig:generated_kernel}) that are used for kernel execution. For CPU backends where the host and device are in the same memory space, some functions (e.g., {\tt cudaMalloc} and  {\tt cudaMemcpy}) can be easily implemented. However, to fully utilize the CPU cores, some effort is required to implement kernel launch and synchronization as well as to handle multi-thread execution. 

For efficient mapping of CUDA kernels to multiple CPU threads, a thread pool is implemented so that only one thread-create and thread-join operation are needed for the entire program. \name{} also implements a task queue to support kernel launches. For each kernel launch, the host thread pushes a kernel variable into the task queue, and the threads inside the thread pool fetch this task from the queue and execute it. Access (push, fetch) to the task queue should be \emph{atomic} to avoid race conditions, so \name{} uses a mutex lock to make the access to the queue atomic.
Listing~\ref{code:kernel_structure} shows the definition of a kernel structure.

%And the synchronization is implemented using a condition variable to block the host thread until the task queue is empty.
%The process of launching and executing a CUDA kernel is shown in Figure~\ref{fig:kernel_launch} with an example kernel {\tt K} with grid size {\tt 2}. 
\begin{lstlisting}[caption={Kernel structure.},label={code:kernel_structure},language=C]
typedef struct kernel {
  void *(*start_routine)(void *);
  void **args;
  dim3 gridDim;
  dim3 blockDim;
  size_t dynamic_shared_mem_size;
  // variables for task queue execution
  int totalBlocks;
  int curr_blockId;
  // used for coarse grain fetching
  int block_per_fetch;
};
\end{lstlisting}

\begin{figure}
    \begin{subfigure}[t]{0.23\textwidth}
      \includegraphics[width=\textwidth]{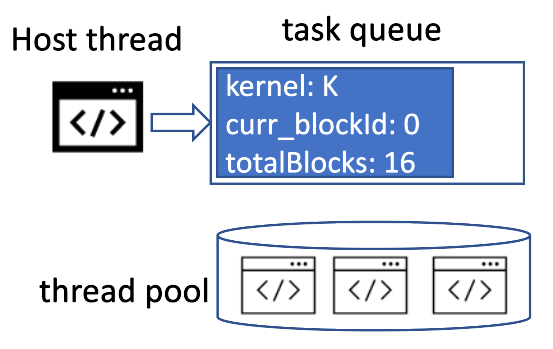}
      \caption{Push.}
    \end{subfigure}
    %\hfill
    \begin{subfigure}[t]{0.23\textwidth}
      \includegraphics[width=\textwidth]{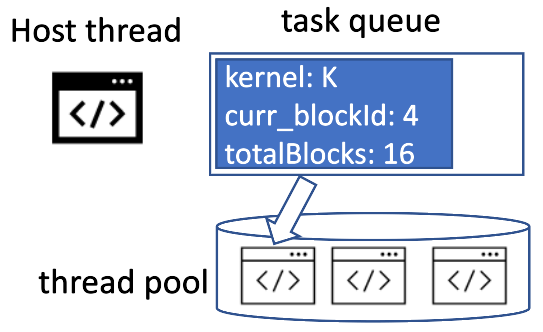}
      \caption{Fetch.}
    \end{subfigure}
    %\hfill
    %\begin{subfigure}[t]{0.23\textwidth}
    %  \includegraphics[width=\textwidth]{launch_2.png}
    %  \caption{Fetch from the queue.}
    %\end{subfigure}
    %\hfill
    %\begin{subfigure}[t]{0.23\textwidth}
    %  \includegraphics[width=\textwidth]{launch_3.png}
    %  \caption{Another idle thread fetches the task and decreases the  grid size. When the grid size becomes zero, the task gets removed from the queue.}
    %\end{subfigure}
    %\hfill
    \caption{The process of launching a CUDA kernel {\tt K} with grid size {\tt 16}.}
    \label{fig:kernel_launch}
  \end{figure}

%, which involves the task push and pop to/from the task queue. The access to the task queue should be \emph{atomic} to avoid race condition. Thus, \name{} uses a mutex lock to make the access to the queue atomic. If there exist a large number of tasks, accessing to the task queue may become the bottleneck.
The process of launching and executing a CUDA kernel is shown in Figure~\ref{fig:kernel_launch} with an example {\tt K} kernel,  {\tt 16} grid size. 
%In Section~\ref{sec:opt}, we discuss two optimizations to alleviate the bottleneck.

\subsubsection{Push task queue (Figure~\ref{fig:kernel_launch}(a))}
For each kernel launch, the host thread first initializes a new kernel structure. As shown in Listing~\ref{code:kernel_structure}, the function pointer variable {\tt start\_routine} points to the kernel function generated by the compilation process, and {\tt args} points to the compressed arguments generated by prologues. The runtime parameters provided at kernel launch are stored in {\tt gridDim}, {\tt blockDim}, and {\tt dynamic\_shared\_mem\_size} variables. In Figure~\ref{fig:kernel_launch}(a), two variables are highlighted. The {\tt curr\_blockId} indicates how many blocks in that kernel have been executed, and the {\tt totalBlocks} indicates how many blocks should be executed for that kernel. \\
After creating the new kernel variable, the host thread pushes it into the task queue. Finally, the host thread broadcasts the condition variable {\tt wake\_pool} to notify the thread pool about the update. Note that the host function is not blocked after kernel launch; after broadcasting the condition signal, it continues to execute subsequent instructions unless it is blocked by a synchronization instruction.

\subsubsection{Fetch task queue (Figure~\ref{fig:kernel_launch}(b))}
At the beginning of execution, all threads inside the thread pool are pending on the condition variable {\tt wake\_pool}. Once the threads are awakened by the broadcast from the host thread, they check the task queue for available tasks. When there are tasks inside the queue, idle threads try to lock the mutex and then fetch the task at the front of the queue. If successful in locking the mutex, the thread updates the attribution of the task: increasing the {\tt curr\_blockId} by {\tt block\_per\_fetch} (4 in the example). {\tt block\_per\_fetch} is a hyperparameter set by \name{} at runtime (see Section~\ref{sec:opt_theory}), and it indicates how many blocks should be executed for each fetch. After the update, if {\tt curr\_blockId} of the task is equal to {\tt totalBlocks}, the task will be popped from the task queue. Finally, the thread unlocks the mutex to let other threads access the task queue. Note that executing a kernel itself is not part of the fetching process, as fetching instructions need to be done atomically and is on the critical path. Separating kernel execution from fetching can increase performance.

\begin{lstlisting}[caption={Set runtime parameters and execute kernels.},label={code:execute_kernel},language=C]
// block_index, block_size and dynamic_shared_memory are thread local variables
for (int s=0;s<blocks_per_fetch;s++){
    block_index = blockId + s;
    block_size = totalBlocks;
    dynamic_shared_memory = (int *)malloc(dynamic_shared_mem_size);
    ker.start_routine(ker.args);
}
\end{lstlisting}

\subsubsection{Execute fetched kernel}
After successfully fetching a task and unlocking the mutex, the thread executes the kernel with the block id from {\tt curr\_blockId} to {\tt curr\_blockId + block\_per\_fetch - 1}. Before executing the transformed kernel function ({\tt start\_routine}), \name{} sets the runtime parameters (e.g., {\tt block\_index, block\_size}) that are needed for execution, as shown in Listing~\ref{code:execute_kernel}.  

% The execution part is shown in 
%In Section~\ref{sec:transform_nvvm_ir}, the kernel function gets the block size from a thread local variable {\tt block\_size}. Also, the dynamic shared memory variable is replaced by {\tt dynamic\_shared\_memory}. Thus, before executing the kernel function, these hyperparameter have to be set (Listing~\ref{code:execute_kernel}).

\ignore{
\subsection{Synchronization}
\ignore{
How to implement multi-thread and how to synchronization?

Host synchronization is achieved by pthread_cond_wait which will wait for the signal from thread pool that all work is done.
}
}

\subsection{Runtime Optimizations\label{sec:opt_theory}}
\subsubsection{Average coarse-grained fetching}
As all accesses to the task queue are atomic, task fetching has non-negligible overhead. Therefore, we implement average coarse-grained fetching where the thread executes $\lceil\frac{gridSize}{threadPoolSize}\rceil$ blocks per fetch (see for-loop in Listing~\ref{code:execute_kernel} line2). This reduces the number of atomic fetches to $threadPoolSize$ while achieving 100\% CPU utilizations, as it equally distributes workloads to all threads inside the pool.
\subsubsection{Aggressive coarse-grained fetching}
\ignore{
Ocelot~\cite{diamos2010ocelot} also implements coarse grain fetching. The {\tt blocks\_per\_fetch} is always set to $\lceil\frac{gridSize}{threadPoolSize}\rceil$, which equally distributes the workload to all the threads. This mechanism can reduce the fetch step to the minimal value and still utilizes all CPU cores. 
}
%If we set {\tt blocks\_per\_fetch} to $\lceil\frac{gridSize}{threadPoolSize}\rceil$, we equally distribute the number of tasks to each thread. This makes the number of fetches to be \emph{minimum} when we make use of all CPU cores. 
We find in some cases, however, that being more aggressive (executing more blocks per fetch) can lead to higher performance. For example, if a CUDA kernel does not have much computation and very few memory accesses, the execution time of a single thread block is extremely short. In this case, we observe that atomic fetching and synchronization among thread pools can be the major performance bottleneck compared to the block execution time. With more aggressive coarse-grained fetching, some threads in the pool stay idle, but the overall time for atomic operations and synchronization is reduced, thereby leading to higher performance in the end.

\begin{figure}
    \begin{subfigure}[t]{0.23\textwidth}
    \centering
      \includegraphics[width=\textwidth]{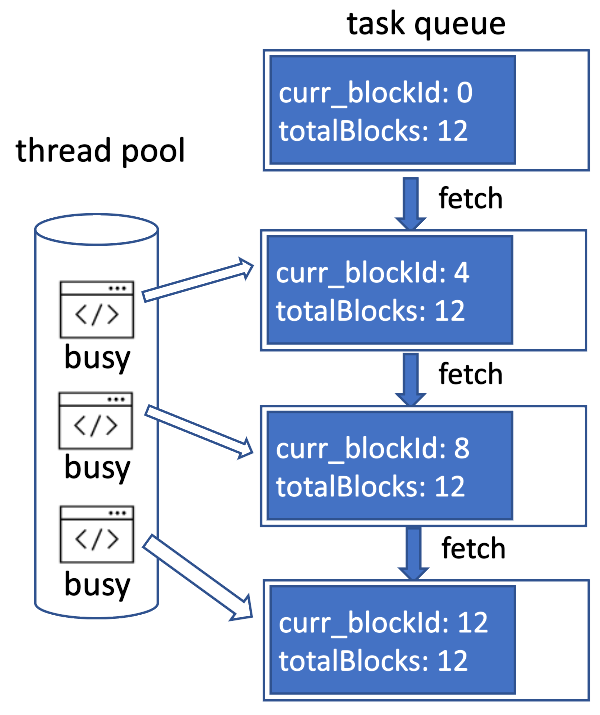}
      \caption{Average coarse-grained fetching.}
    \end{subfigure}
    \begin{subfigure}[t]{0.23\textwidth}
    \centering
      \includegraphics[width=\textwidth]{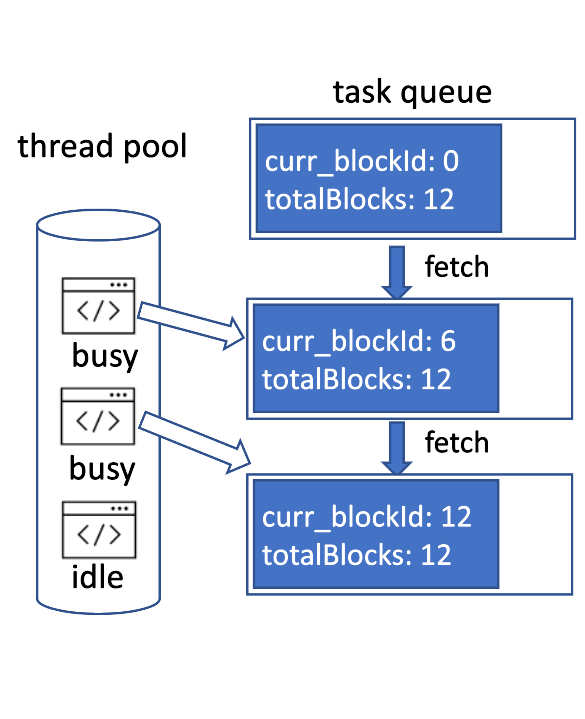}
      \caption{Aggressive coarse-grained fetching.}
    \end{subfigure}
    \caption{The process of executing CUDA kernels (grid size=12) with the thread pool size of 3.}
    \label{fig:runtime_opt}
  \end{figure}

Figure~\ref{fig:runtime_opt} shows the difference between average and aggressive coarse-grained fetching. Assume that we have a CUDA kernel with a grid size of 12, and there are three threads in the thread pool. With average coarse-grained fetching (Figure~\ref{fig:runtime_opt}(a)), four blocks are executed for every atomic fetch. Thus, we need three fetches while all threads within the pool can be utilized. In Figure~\ref{fig:runtime_opt}(b), with aggressive coarse-grained fetching, six blocks are fetched in each step, so we end up performing two atomic fetches. In this case, only two threads are utilized, and the other one thread is idle during the execution. However, the overall execution time is smallest with aggressive fetching if the CUDA kernel is short. Since there is a trade-off between CPU utilization and reducing the number of atomic fetch operations, \name requires several heuristics to find the optimal fetching block size (aka grain size) per kernel. Section~\ref{sec:aggressive_evaluate} discusses these heuristics and the performance results.

%lthough both of the optimizations above perform coarse-grained fetching, the first one always results in improving performance as it decreases the overhead of atomic fetching without any side effect. For the second one, however, we may not make use of all available CPU (hardware) threads. Thus, there is a trade-off between the CPU utilization and the number of atomic fetching. 
%
%The first one also always packs the $\lceil\frac{grid size}{threadPoolSize}\rceil$ number of GPU blocks together for a coarse-grained task, while the second one needs to manually select the number of GPU blocks packed together. 
%
%Section~\ref{sec:aggressive_evaluate} discusses the performance results and analysis for the optimizations.

\section{Evaluation}
In the following two sections, we try to answer the following questions with the experiments:
\begin{itemize}
    \item Can \name{} achieve higher coverage of common CUDA kernels compared with source-to-source translators? (Section~\ref{sec:coverage})
    \item What is the end-to-end performance of our approach \name{}, and other CUDA-on-CPU projects? (Section~\ref{sec:performance})
    \item What is the end-to-end performance of \name{} and manually migrated CPU code using OpenMP? (Section~\ref{sec:cloverleaf_openmp})
    \item What is the kernel performance difference between CPU and GPU, and can CPUs effectively run CUDA kernels with \name{}? (Section~\ref{sec:future})
\end{itemize}
%In this section, \name{} is evaluated on several CUDA benchmarks on three ISAs: x86 (Intel CPU, AMD CPU), AArch64 (Arm CPU) and RISC-V (SiFive CPU). 
To answer the first two questions, two frameworks, DPC++~\cite{reinders2021data} and HIP-CPU~\cite{HIP-CPU}, are selected as the baseline. 
%Although there are other projects also support executing CUDA on CPU, some of them has not been maintained long ago~\cite{stratton2008mcuda,diamos2010ocelot}, and some are proposed for debugging CUDA programs on CPU~\cite{blomkvist2021cumulus,elhelw2020horus}, not for performance.
Six servers are used as the hardware platform (Table~\ref{table:amd_cuda_conf}). %Server-CUDA is configured with an NVIDIA Ampere GPU, as the baseline. Server-Intel, Server-ARM and Server-SiFive are configured with Intel, ARM, and SiFive CPU separately. 
Detailed information of software versions and input size for each benchmark are reported in the Artifact Description.

\subsection{Coverage\label{sec:coverage}}

\subsubsection{Architecture coverage among different platforms}
Table~\ref{table:hardware_coverage} summarizes the architecture coverage of different platforms. 

\begin{table}[h!]
\centering
\resizebox{\columnwidth}{!}{%
\begin{tabular}{|c|c|c|c|}
\hline
Framework & \begin{tabular}[c]{@{}c@{}}Compilation\\ requirement\end{tabular} & \begin{tabular}[c]{@{}c@{}}Runtime\\ requirement\end{tabular} & \begin{tabular}[c]{@{}c@{}}ISA \\ support\end{tabular}         \\ \hline \hline
DPC++     & DPC++                                                             & DPC++                                                         & x86                                                            \\ \hline
HIP-CPU   & C++17                                                             & TBB(\textgreater{}=2020.1-2), pthreads                        & \begin{tabular}[c]{@{}c@{}}x86\\ AArch64\\ RISC-V\end{tabular} \\ \hline
CuPBoP    & LLVM                                                              & pthreads                                                      & \begin{tabular}[c]{@{}c@{}}x86\\ AArch64\\ RISC-V\end{tabular} \\ \hline
\end{tabular}%
}
\caption{Comparisons among DPC++,  HIP-CPU, and \name{}.}
\label{table:hardware_coverage}
\end{table}
DPC++ is a SYCL~\cite{howes2015sycl} implementation developed by Intel, and DPCT~\cite{DPCT} is a provided tool that supports translating CUDA source code to SYCL source code. DPC++ supports Intel CPU, GPU, and FPGA backends with some manual code optimization required~\cite{one-api-requirement}.%\footnote{https://www.intel.com/content/www/us/en/developer/articles/system-requirements/intel-oneapi-dpcpp-system-requirements.html}

HIP~\cite{HIP} is a programming language developed by AMD that has been developed for programming on AMD GPUs. The AMD team provides the Clang-based HIPIFY~\cite{HIPIFY} tool to support translating CUDA source code to HIP source code. HIP-CPU is a header file library that implements HIP runtime functions (e.g., {\tt hipLaunchKernelGGL}, {\tt hipMemcpy}) on CPUs, so it can be deployed on different CPU devices as long as all its dependency libraries (like TBB 2020.1) can be built. %HIP-CPU dependents on C++17 and the latest TBB (>=2020.1), which makes it difficult to be supported on old machines. 

As mentioned earlier, \name{} is based on LLVM and utilizes both LLVM and NVVM IR to generate binary files for different backend devices. This means that code compiled with \name can run on any supported LLVM CPU backends, including x86, AArch64, and RISC-V. 

%LLVM is only required for \name{} compilation. As \name{} has separate compilation and runtime, it supports cross-compile. Users can use devices that contains LLVM to build executable files for other devices. We have verified the cross-compile features by compiling on x86 host machine and executing on Arm/RISC-V devices. 
%The hardware coverage and dependency information are concluded in Table~\ref{table:hardware_coverage}.

\subsubsection{Benchmark Coverage}
We used the Rodinia~\cite{che2009rodinia}, Crystal~\cite{shanbhag2020crystal}, and Hetero-Mark benchmarks~\cite{sun2016hetero} to evaluate the coverage. For the Rodinia benchmark, an already transformed code repository exists for DPC++~\cite{castano2021intel}, and for HIP-CPU we use HIPIFY to translate Rodinia CUDA code to HIP. Table~\ref{table:rodinia_coverage} shows the coverage. \par
For Rodinia benchmark, as none of these frameworks support texture memory on CPUs, hybridsort, kmeans, leukocyte, and mummergpu examples are not supported. HIP-CPU also does not support c-based codes like b+tree and backprop.  For the dwt2d example, the generated NVVM IR uses several NVIDIA-specific intrinsic functions (e.g., {\tt \_\_nvvm\_d2i\_lo, \_\_nvvm\_lohi\_i2d}) for which NVIDIA does not provide the details; thus, \name{} cannot support it currently. Overall \name{} achieves the highest coverage 69.6(\%). We look to support texture memory in future work. \par
The Crystal benchmark is a high-performance GPU database implementation that contains SQL query operator implementations and implementations of 13 database queries. \name{} supports all 13 queries (100\%). HIP-CPU doesn't support CUDA warp shuffle, so it only supports 10 queries (76.9\%). As DPC++ doesn't support atomicCAS on CPU devices, it cannot support any of the Crystal queries. \par
Hetero-Mark's examples are relatively simple and do not involve complex grammar and CUDA features. As a result, all three frameworks support eight out of 10 benchmarks. BST and KNN rely on CUDA system-wide atomics features that are not supported by these frameworks. We were not able to run BE, as it relies on OpenCV and was not available in all test environments.
\ignore{
For source-to-source translation, due to the flexible of high-level languages, the translator can hardly deal with all the cases. Thus, manually modification maybe required for either input (CUDA source code) or output (translated high-level languages). In this section, we analyzes the manually modification required for DPC++, HIP-CPU, and \name{}. Compared with Rodinia benchmark, Hetero-mark benchmark has relatively simple programs, and all three framework can successfully translate and execute these examples. Thus, we only discuss Rodinia benchmark. 
}

\begin{table}[]
\centering
\resizebox{\columnwidth}{!}{%
\begin{tabular}{|c|c|c|c|c|}
\hline
Name                                                                               & DPC++                            & HIP-CPU                          & CuPBoP                         & features                                                                   \\ \hline
b+tree                                                                             & correct                          & unsupport                        & correct                        & extern C                                                                   \\ \hline
backprop                                                                           & correct                          & unsupport                        & correct                        & extern C                                                                   \\ \hline
bfs                                                                                & incorrect                        & correct                          & correct                        &                                                                            \\ \hline
gaussian                                                                           & correct                          & correct                          & correct                        &                                                                            \\ \hline
hotspot                                                                            & incorrect                        & correct                          & correct                        &                                                                            \\ \hline
hotspot3D                                                                          & incorrect                        & correct                          & correct                        &                                                                            \\ \hline
huffman                                                                            & correct                          & unsupport                        & correct                        & \begin{tabular}[c]{@{}c@{}}extern shared \\ memory definition\end{tabular} \\ \hline
lud                                                                                & correct                          & correct                          & correct                        &                                                                            \\ \hline
myocyte                                                                            & correct                          & correct                          & correct                        &                                                                            \\ \hline
nn                                                                                 & correct                          & correct                          & correct                        &                                                                            \\ \hline
nw                                                                                 & correct                          & correct                          & correct                        &                                                                            \\ \hline
particlefilter                                                                     & incorrect                        & correct                          & correct                        &                                                                            \\ \hline
pathfinder                                                                         & correct                          & correct                          & correct                        &                                                                            \\ \hline
srad                                                                               & correct                          & correct                          & correct                        &                                                                            \\ \hline
streamcluster                                                                      & correct                          & correct                          & correct                        &                                                                            \\ \hline
dwt2d                                                                              & segfault                         & unsupport                        & unsupport                      & \begin{tabular}[c]{@{}c@{}}shared memory\\  for structure\end{tabular}     \\ \hline
hybridsort                                                                         & unsupport                        & unsupport                        & unsupport                      & Texture                                                                    \\ \hline
kmeans                                                                             & unsupport                        & unsupport                        & unsupport                      & Texture                                                                    \\ \hline
lavaMD                                                                             & correct                          & correct                          & unsupport                      & intrinsic function                                                         \\ \hline
leukocyte                                                                          & unsupport                        & unsupport                        & unsupport                      & Texture                                                                    \\ \hline
mummergpu                                                                          & unsupport                        & unsupport                        & unsupport                      & Texture                                                                    \\ \hline
cfd                                                                                & correct                          & unsupport                        & correct                        & cuGetErrorName                                                             \\ \hline
heartwall                                                                          & incorrect                        & unsupport                        & incorrect                      & complex template                                                           \\ \hline
\begin{tabular}[c]{@{}c@{}}Rodinia\\ coverage\end{tabular}                         & 56.5                             & 56.5                             & 69.6                           &                                                                            \\ \hline
{ q11, q12, q13}                                               & { unsupport} & { unsupport} & { support} & { warp shuffle}                                        \\ \hline
{ q21, q22, q23}                                               & { unsupport} & { support}   & { support} & {atomicCAS}                                           \\ \hline
{ q31, q32, q33, q34}                                          & { unsupport} & { support}   & {support} & {atomicCAS}                                           \\ \hline
{q41, q42, q43}                                               & {unsupport} & { support}   & {support} & {atomicCAS}                                           \\ \hline
{\begin{tabular}[c]{@{}c@{}}Crystal\\  coverage\end{tabular}} & {0}         & { 76.9}      & {100}     & { }                                                    \\ \hline
\end{tabular}%
}
\caption{Platform coverage results of Rodinia (applications from b+tree to heartwall) and Crystal (applications from q11 to q43) benchmarks.}
\label{table:rodinia_coverage}
\end{table}

\subsubsection{HPC application support}
As an example case of HPC applications, we migrate CloverLeaf-CUDA to CPUs. CloverLeaf~\cite{mallinson2013cloverleaf} is a mini-app that solves the compressible Euler equations. The CUDA implementation contains 18 CUDA kernels. The host programs contain both C++ and Fortran. \name{} can support the transformation of CloverLeaf-CUDA. However, both DPCT and HIPIFY fail to translate macros such as the one shown in Listing~\ref{code:complex_macro}. 

\begin{lstlisting}[caption={Complex macro that cannot be translated by DPCT or HIPIFY.},label={code:complex_macro},language=C]
#define CUDALAUNCH(funcname, ...)  \
    funcname<<<num_blocks, BLOCK_SZ>>>(x_min, x_max, y_min, y_max, __VA_ARGS__);
\end{lstlisting}

From the above evaluation, we can conclude why ~\name{} achieves higher coverage. As both DPCT and HIPIFY rely on source-to-source translation, they need to generate human-readable code as output. Thus, they will not expand macros during translation to avoid generating complex code. As \name{} implements transformations on the LLVM IR level, it doesn't need to handle these macros. 
Another challenge for source-to-source translation comes from differences in APIs. For example, DPC++ uses exceptions to raise errors, while CUDA uses error codes~\cite{castano2021intel}.
%This evaluation demonstrates that both DPCT and HIPFY need manual modifications on the generated code in large CUDA applications that rely on specific C++ language features. Since \name{} shares the same front-end (clang) as CUDA, it avoids this problem.

%to implement CUDA kernel launch. Neither DPCT nor HIPIFY can correctly translate this macro.

%CUDA is based on C++, which is too flexible to support all the cases. Thus, DPCT and HIPIFY cannot be easily used for real applications. And manually modifications for the generated code are always required. Instead of translating CUDA source language, \name{} compiles CUDA to LLVM IR and applies transformation on these IRs.
%Also, \name{} directly maintains a runtime library which implements functions have the same signature as CUDA. In a word, to support a new CUDA feature, \name{} only needs to maintain the runtime library, while DPC++ and HIP-CPU needs to modify both the source-to-source translator and runtime library.

\subsection{Performance\label{sec:performance}}
 In this section, for all performance comparisons, we measure the end-to-end performance of the entire kernel or application execution, including the data transfer time to/from the GPU. %Thereby, depending on the size of data to transfer, there is advantage of using CPUs only rather than CPU+GPU~\cite{qilin}. 

\subsubsection{Performance on x86}
Table~\ref{table:intel_performance} shows the end-to-end execution time: CUDA is executed on Server-Intel-GTX and the rest are executed on Server-Intel.\footnote{ The configuration of each benchmark samples is recorded in the Artifact Description.}

We also provide the OpenMP execution for Rodinia as a reference for ``native CPU'' execution but note that detailed comparisons with OpenMP are outside the scope of this work. Overall, \name{} achieves comparable performance with DPC++ and HIP-CPU. Compared with CUDA, on average, DPC++ is 83\% slower, HIP-CPU is 156\% slower, and \name{} is 96\% slower. These results also demonstrate that end-to-end time, including the data transfer between host and device, can provide motivation for executing code on CPUs rather than on CPUs and GPUs with limited PCIe bandwidth~\cite{qilin}.
%The performance gap between CPUs and GPUs are much smaller than the peak performance gap.  

\begin{table*}
\centering
\tiny
\resizebox{2.0\columnwidth}{!}{%
\begin{tabular}{|c|c|c|c|c|c|c|}
\hline
\multirow{2}{*}{Server Name}      & \multirow{2}{*}{CPU / GPU} & \multirow{2}{*}{CPU cores / GPU SMs} & \multirow{2}{*}{Peak FLOP/s} & \multirow{2}{*}{Memory (GB)} & \multirow{2}{*}{Peak Memory BW (GB/s)} & \multirow{2}{*}{L2 cache / shared memory} \\
                                  &                            &                                      &                              &                              &                                        &                                           \\ \hline \hline
Server-Intel                      & Intel Gold6226R (x2)       & 32                                    & 972G                         & 376                          & 140                                    & 16 MB                                     \\ \hline
\multirow{2}{*}{Server-AMD-A30}   & AMD EPYC 7502 (x2)         & 64                                   & 123G                         & 264                          & 409.6                                  & 16 MB                                     \\ \cline{2-7} 
                                  & NVIDIA A30 GPU             & 56                                  & 10.3T                        & 24                           & 933                                    & 128KB                                    \\ \hline
\multirow{2}{*}{Server-Intel-GTX} & Intel i7-11700             & 8                                    & 200G*                         & 32                           & 50                                     & 2 MB                                     \\ \cline{2-7} 
                                  & GTX 1660Ti                 & 24                                   & 5.4T                         & 6                            & 288                                    & 32KB                                \\ \hline
Server-Arm1                       & Arm A64FX                  & 48                                   & 2.7T                         & 32                           & 1024                                   & 8 MB                                     \\ \hline
Server-Arm2                       & Arm Altra Q80-30           & 80                                   & 3.8T                            & 512                          & 102.4*                                  & 1 MB                                     \\ \hline
Server-SiFive                     & SiFive FU740 (U74)         & 4                                    & -                            & 16                           & -                                      & 128 KB                                    \\ \hline
\end{tabular}%
}
\caption{Hardware environment configurations. *estimated value}
\label{table:amd_cuda_conf}
\end{table*}

\begin{table}[]
\centering
\tiny
\begin{tabular}{|c||c|c|c|c|c|c|}
\hline
benchmark                    & samples        & CUDA  & DPC++  & HIP       & \name{} & OpenMP \\ \hline \hline
\multirow{14}{*}{Rodinia}    & b+tree         & 1.459 & 1.577  & unsupport & 2.135  & 1.56   \\ \cline{2-7} 
                             & backprop       & 0.672 & 2.51   & unsupport & 1.964  &        \\ \cline{2-7} 
                             & bfs            & 1.29  & 1.555  & 1.267     & 1.136  & 1.365  \\ \cline{2-7} 
                             & gaussian       & 0.866 & 1.12   & 8.494     & 1.669  &        \\ \cline{2-7} 
                             & hotspot        & 1.239 & 1.373  & 1.267     & 1.072  & 1.11   \\ \cline{2-7} 
                             & hotspot3D      & 1.376 & 1.249  & 1.732     & 1.269  & 1.262  \\ \cline{2-7} 
                             & lud            & 0.68  & 1.212  & 0.953     & 1.164  & 0.082  \\ \cline{2-7} 
                             & myocyte        & 1.087 & 3.327  & 0.397     & 0.151  &        \\ \cline{2-7} 
                             & nn             & 0.443 & 2.004   & 1.198    & 1.309  &        \\ \cline{2-7} 
                             & nw             & 1.068 & 2.126  & 1.767     & 1.589  & 0.477  \\ \cline{2-7} 
                             & particlefilter & 0.751 & 0.889  & 0.836     & 0.833  & 0.702  \\ \cline{2-7} 
                             & pathfinder     & 1.92  & 2.395  & 2.424     & 2.359  &        \\ \cline{2-7} 
                             & srad           & 1.979 & 5.996  & 8.308     & 2.886  & 2.474  \\ \cline{2-7} 
                             & streamcluster  & 6.607 & 14.804 & 21.09     & 18.435 & 13.977 \\ \hline
\multirow{8}{*}{Hetero-Mark} & AES            & 29.87 & 48.381 & 55.595    & 50.107 &        \\ \cline{2-7} 
                             & BS             & 0.967 & 1.504  & 2.506     & 2.74   &        \\ \cline{2-7} 
                             & ep             & 4.187 & 2.506  & 34.085    & 28.844 &        \\ \cline{2-7} 
                             & fir            & 1.445 & 4.389  & 4.225     & 3.872  &        \\ \cline{2-7} 
                             & ga             & 0.846 & 1.598  & 2.256     & 1.959  &        \\ \cline{2-7} 
                             & hist           & 1.829 & 2.529  & 2.309     & 2.78   &        \\ \cline{2-7} 
                             & kmeans         & 2.968 & 1.513  & 4.581     & 5.165  &        \\ \cline{2-7} 
                             & PR             & 2.836 & 3.506  & 3.789     & 4.783  &        \\ \hline
\end{tabular}
\caption{End-to-end execution time (sec) for Rodinia and Hetero-Mark benchmarks in Server-Intel-GTX (CUDA) and Server-Intel (DPC++, \name{}, HIP-CPU).}
\label{table:intel_performance}
\end{table}

\ignore{
  \begin{figure*}[h!]
    \centering
    \includegraphics[width=170mm]{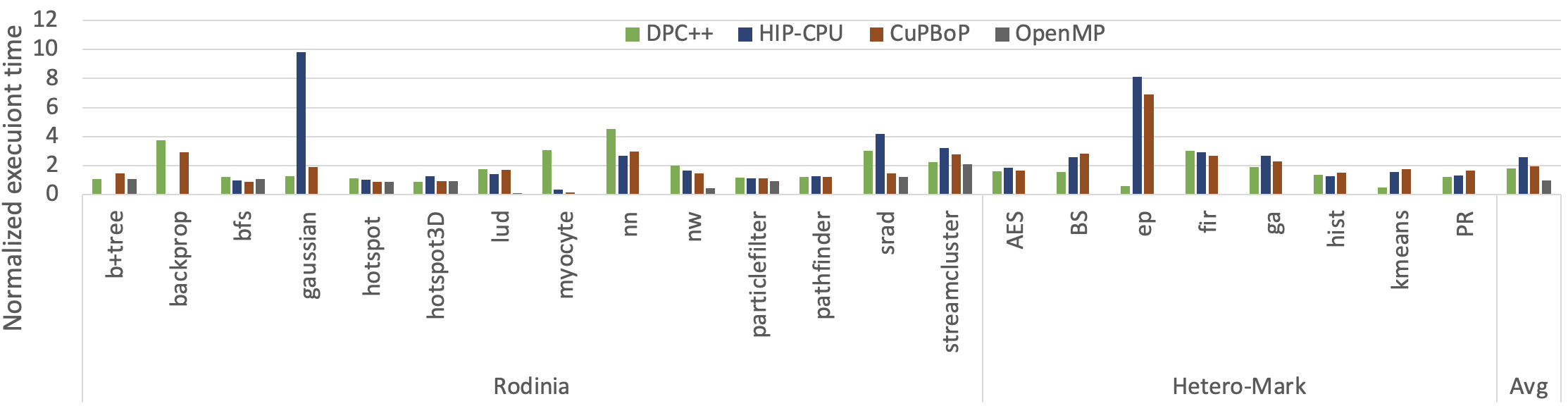}
    \caption{The end-to-end execution time normalized with CUDA execution time. CUDA is evaluated on Server-Intel-GTX. DPC++, \name{}, and HIP-CPU are evaluated on Server-Intel.}
    \label{fig:normalized_intel}
  \end{figure*} 
 }
  
% As OpenMP doesn't have unnecessary memory movement (host-device communication) and be optimized with SIMD pragma, in some cases the end2end time is even faster than CUDA on GPU. 
OpenMP kernels have different code structures, as the OpenMP API and runtime are natively targeted to CPUs and rely on compiler support for proper vectorization and parallelization. Thus, OpenMP execution time has high variance compared with the other platforms with some small kernels running faster with OpenMP than CUDA because the kernels' small grid size cannot fully utilize the computation resources of CUDA. Thus, comparing with OpenMP is not the major concern in this work. 

DPC++ and \name{} have similar performance, while HIP-CPU is much slower for the tested kernels. However, we note that several kernels have large variations in performance between the three frameworks and OpenMP execution. \\
\begin{lstlisting}[caption={CUDA code that can be optimized by vectorization.},label={code:cuda_vectorization},language=C]
// code from EP
for (int j = 0; j < num_vars; j++) {
  double pow = 1;
  for (int k = 0; k < j + 1; k++) {
    pow *= creature.parameters[j];
  }
  fitness += pow * fitness_function[j];
}
// code from KMeans
for (int i = 0; i < nclusters; i++) {
    float dist = 0;
    float ans = 0;
    for (int l = 0; l < nfeatures; l++) {
        ans += (feature[l * npoints + point_id] - clusters[i * nfeatures + l]) * (feature[l * npoints + point_id] - clusters[i * nfeatures + l]);
      }
    dist = a
    if (dist < min_dist) {
        min_dist = dist;
        index = i;
    }
}
// code from NW
int bx = blockIdx.x;
int tx = threadIdx.x;
int b_index_x = bx;
int b_index_y = i - 1 - bx; // i is input
int index_w   = cols * BLOCK_SIZE * b_index_y + BLOCK_SIZE * b_index_x + ( cols );
temp[tx + 1][0] = matrix_cuda[index_w + cols * tx];
\end{lstlisting}

\textbf{Gaussian:} In the Gaussian, the kernel invokes a large number of blocks (65536). As mentioned in Section \ref{sec:opt_theory}, a large number of blocks will significantly increase the time for fetching the task queue. Without coarse-grained fetching, \name{} needs 82.277 sec to complete the program. HIP-CPU doesn't do special optimization on this situation, so it has the longest execution time. \\
\textbf{Myocyte:} The host program launches a kernel 3781 times, with grid size equal to 2 and block size equal to 32. Most instructions in the launched kernel are compute-intensive with very little memory access. With such a lightweight workload, the kernel launch time becomes the bottleneck that can be improved using aggressive coarse-grain fetching. Before the optimization, \name{} takes 1.659 sec. After aggressive coarse-grain fetching, it only requires 0.151 sec. Launching many small kernels also hurts the performance of CUDA, and all three tested frameworks and CUDA are much slower than the OpenMP code which only contains SIMD loops. \\
\textbf{Srad:} Similar to the Gaussian, it also launches a large grid kernel with grid size 262144. In that case, coarse-grain fetching is needed to avoid the fetching task queue becoming the bottleneck. There are also nine block synchronizations ({\tt \_\_syncthreads}) within the kernels. As HIP-CPU uses fibers instead of a for-loop to implement the threads within a block, it has higher overhead for context switching during block synchronization, which makes it the slowest framework. \\
\textbf{EP:} EP contains a kernel that has a nested loop, as shown in Listing~\ref{code:cuda_vectorization} lines 4-6. DPC++ can vectorize the inner loop while LLVM cannot. Therefore, DPC++ shows significantly better performance than \name{} and HIP-CPU. \\
\textbf{KMeans:} The same as EP, KMeans has a loop that can only be vectorized by DPC++.(Listing~\ref{code:cuda_vectorization} lines 10-21) \\
%contains a kernel which can be accelerate by SIMD instructions for x86 architecture. However, only DPC++ supports auto-vectorizing this kernel.
\textbf{NW:} This example contains a complex index operation to access array elements: Listing~\ref{code:cuda_vectorization} lines 27-28.  Hence, OpenMP was able to vectorize with programmer-based hints, but DPC++, \name{}, and HIP-CPU cannot perform compiler-driven vectorization.  \\

%vectorizated computation and memory access pattern, which can be accelerated by SIMD instruction. However, these patterns are too complex to be optimized by compilers. In OpenMP, the SIMD is implemented by pragma written by programmers. Thus, OpenMP can achieve the highest performance.
%The code that can be optimized by vectorization is listed in Code~\ref{code:cuda_vectorization}.

\subsubsection{Performance on aarch64 and RISC-V}
\begin{figure}[h!]
    \centering
    \includegraphics[width=80mm]{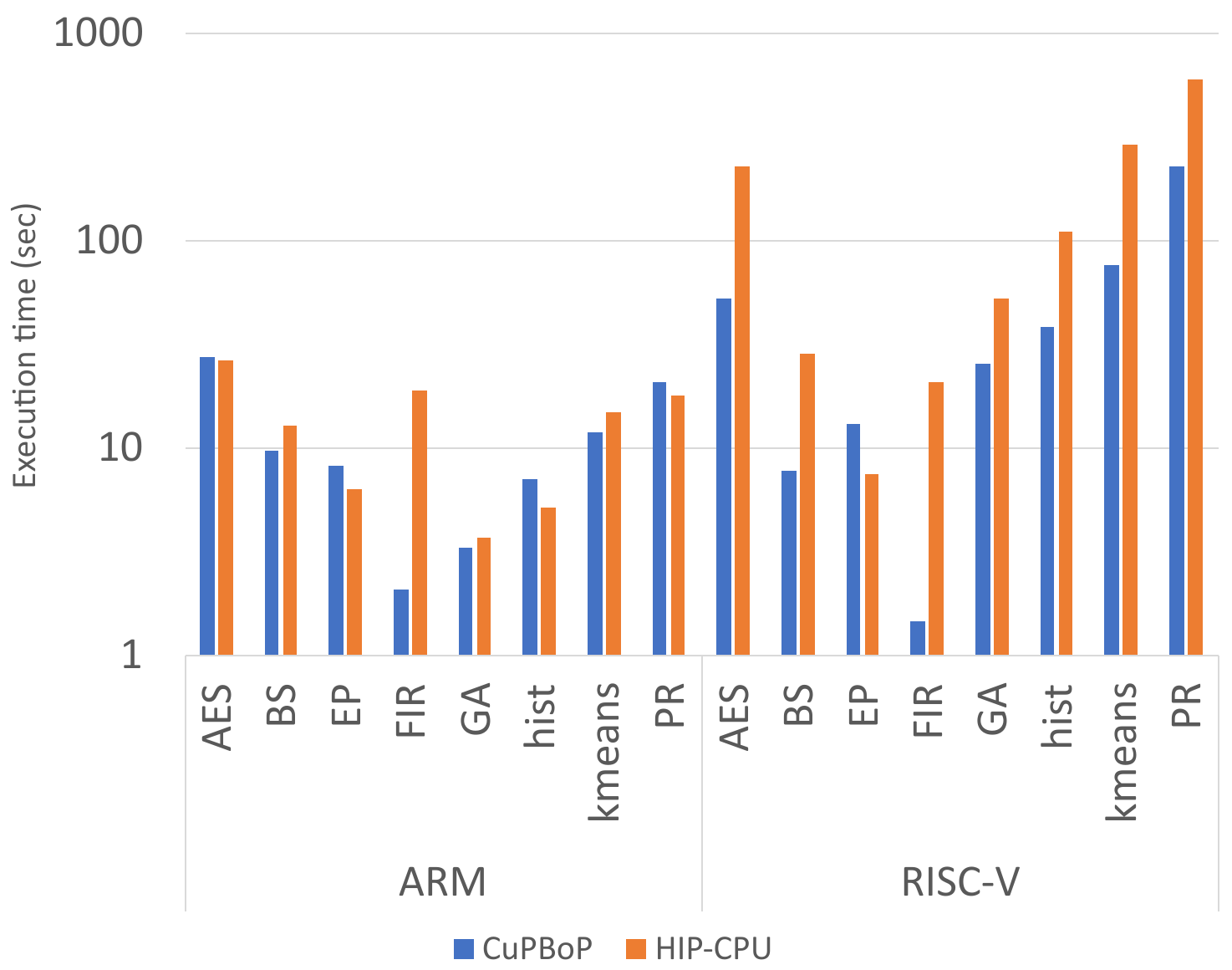}
    \caption{The execution result on Arm and RISC-V devices (Server-Arm1, Server-SiFive). }
    \label{fig:Arm_riscv}
  \end{figure}

Figure~\ref{fig:Arm_riscv} shows the performance of the Hetero-Mark benchmark using \name{} and HIP-CPU with Server-Arm1 and Server-SiFive. In all these benchmarks, \name{} achieves higher performance than HIP-CPU. HIP-CPU has to apply synchronizations before any memory copy between host and device to guarantee the correctness, which can be a performance overhead. For the FIR example that contains a large number of memory copies, many unnecessary synchronizations applied by HIP-CPU significantly hurt the performance. Instead, \name{} only applies synchronizations after kernel launches that write memory addresses that are read by later instructions. On average, \name{} is 30\% faster than HIP-CPU. 

\subsection{Aggressive coarse grain fetching\label{sec:aggressive_evaluate}}

%Aggressive coarse grain is introduced in Section~\ref{sec:opt}. Aggressive coarse grain is a trade-off between CPU utilization and race condition. If we equally distribute the workload to all CPU cores, we have 100\% utilization, however, all CPUs compete the shared variables. In contrast, if use extremely coarse grain which we assign all workloads to a single CPU core, there will be no competition, but the CPU utilization will be low. \\

%In this section, case study is used to analyze the performance gain from this trade-off. We use examples from the Hetero-mark benchmark as it provides knobs to tune the data scale for evaluation. Besides, to avoid the effect of multiple kernels, we only evaluate the examples which contains single kernels. 

Table~\ref{table:opt_evaluate} shows the performance of the Hetero-Mark  when we use different grain sizes (the number of blocks to execute per fetch).\footnote{ To simplify the evaluation, we use benchmarks that have only one kernel.}  Overall, benchmarks that have many fewer instructions per kernel see more benefits from aggressive coarse-grain fetching.The instruction count is collected using the nvprof profiler, and the BS and FIR have many fewer instructions than other tested kernels.  Although HIST has a relatively heavy workload, it can also benefit from aggressive coarse-grain fetching. This is due to HIST containing atomic instructions, and with aggressive coarse-grain fetching there are fewer activated threads competing for shared locks during atomic updates. To confirm this, we intentionally replace atomic instructions with normal store instructions (HIST-no-atomic) and measure the performance. When HIST does not have any atomic instructions inside the kernel, fully utilizing the entire CPU threads (average coarse-grain fetching) shows the best performance. 

%If replacing the atomic instruction with normal store instruction, hist cannot gain performance improvement by using aggressive coarse grain fetching (row 4,5-th in Table~\ref{table:opt_evaluate}). This phenomenon also shown only small kernels can benefit from aggressive coarse grain fetching, without taking atomic instructions into account.\\

%The red color means grain size that equally distributes workloads to all CPU cores, as used in Ocelot~\cite{diamos2010ocelot}. While the green color means the higher performance cases when using aggressive coarse grain fetching. \\
%When CUDA kernels are relatively small, atomically fetching the task queue is the bottleneck for execution. With large coarse grain, although it cannot fully utilize all cores, it reduces the number of atomically fetching. \\
%By using nvprof, we profile the CUDA kernel workloads in these examples (Table~\ref{table:nvprof_prof}. From the profiling results, BS and FIR have much light workloads, which makes them benefit from aggressive coarse grain fetching. One thing worth noting is that although hist has relatively heavy workload, it can also benefit from aggressive coarse grain fetching. This is due to hist example contains atomic instructions. When using aggressive coarse grain fetching, there are fewer activated threads, which means fewer competition during atomic update. \\

\begin{table}[]
\tiny
\centering
\begin{tabular}{|c||c|c|c|c|c|c|c||c|}
\hline
time (s)         & 1                             & 2                            & 4     & 8                            & 16     & 24     & 32    & \# inst  \\ \hline
BS               & {\color[HTML]{FE0000} 8.93}   & 8.302                        & 7.853 & {\color[HTML]{34FF34} 7.766} & 11.452 & 15.318 & 19.216 &  79k \\ \hline
FIR              & {\color[HTML]{FE0000} 10.299} & 9.145                        & 8.424 & {\color[HTML]{34FF34} 8.209} & 8.802  & 10.558 & 12.83 &  260k \\ \hline
GA               & {\color[HTML]{FE0000} 2.341}  & 3.285                        & 4.586 & 7.535                        & 12.383 & 12.593 & 12.637 & 25M \\ \hline
HIST             & 2.813                         & {\color[HTML]{FE0000} 2.733} & 2.579 & {\color[HTML]{34FF34} 2.499} & 3.012  & 3.303  & 3.498  &  15M \\ \hline
HIST (no atomic) & 1.032                         & {\color[HTML]{FE0000} 1.016} & 1.178 & 1.537                        & 1.913  & 2.245  & 2.564  & 15M \\ \hline
PR               & 3.928                         & {\color[HTML]{FE0000} 3.782} & 5.782 & 8.952                        & 12.691 & 15.501 & 17.778 & 14M \\ \hline
AES               & {\color[HTML]{FE0000} 1.927}  & 2.798                        & 5.852 & 9.876                        & 17.38 & 23.864 & 28.851 & 9M\\ \hline
\end{tabular}
\caption{Execution time when using different grain size. The cases shown with red mean equal distributions of workloads to all threads (average coarse-grain fetching). The green numbers mean the best result with aggressive coarse-grain fetching. \# inst shows the number of executed instructions for each kernel. }
\label{table:opt_evaluate}
\end{table}

\section{Discussions: Can \name{} run CUDA faster than NVIDIA GPU?~\label{sec:future}}
\name{} achieves comparable or even higher performance compared with other frameworks that execute CUDA on CPUs. In this section we further investigate the performance differences between \name{} and manually optimized code using roofline models to see how far current high-end CPUs are from the performance peaks of a high-end NVIDIA Ampere GPU. The hardware configuration is shown in Table~\ref{table:amd_cuda_conf}.

\subsection{CloverLeaf analysis\label{sec:cloverleaf_openmp}}
%First, CloverLeaf~\cite{CloverLeaf} mini-application is used for evaluation. We try to use \name{}, DPCT and HIPIFY to translate the CUDA source code. Due to the complex macro cases in the source code, DPCT and HIPIFY generate incorrect DPC++ or HIP code. Only \name{} can generate the correct code. 
CloverLeaf has manually optimized OpenMP and MPI implementations, and \name{} is compared with them on different servers in Figure~\ref{fig:CloverLeaf} (CUDA baseline is executed on Server-AMD-A30). \name{} is slower than both the OpenMP and MPI implementations, which indicates that \name{} does not currently achieve the peak performance on CPU architectures. In the next section, we use roofline models and CPU cache analysis to study the performance gap between \name{} and the CPU peak performance.
\ignore{
Profiling with Intel SDE~\cite{IntelSDE}, we get the following results (Table~\ref{table:cloverleaf_prof}), which indicates CuPBoP has low utilization for SIMD instructions and poor cache locality. For Mem access, due to the low cache hit rate, \name{} has to f
\begin{table}[]
\centering
\begin{tabular}{|c|c|c|c|}
\hline
Metrics         & CuPBoP     & OpenMP     & MPI        \\ \hline
Mem access (GB) & 4067       & 28         & 193        \\ \hline
\# of inst      & 7.8 * 1e12 & 8.4 * 1e10 & 4.1 * 1e11 \\ \hline
\end{tabular}
\caption{CloverLeaf profiling results for different frameworks.}
\label{table:cloverleaf_prof}
\end{table}
}
%This is due to \name{} is directly tuned from CUDA, which may contains memory access pattern that inefficient for CPU cache system. Besides, \name{} cannot utilize SIMD instruction efficiently, as the transformed LLVM IRs are too complex to be auto-vectorized by existing LLVM. 
  \begin{figure}[htbp]
    \centering
    \small
    \includegraphics[width=80mm]{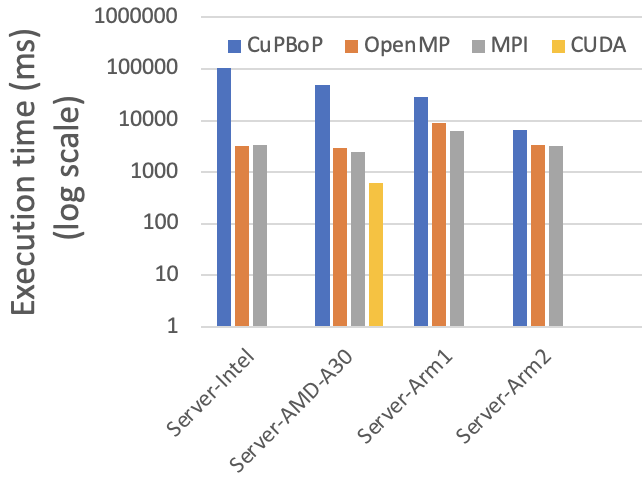}
    \caption{The end2end execution time for CloverLeaf.}
    \label{fig:CloverLeaf}
  \end{figure} 
  
%\name{} is executed on 64 cores/128 threads CPU (released in 2019) with CUDA 
%In this section, several cases study sampled from Rodinia and Hetero-mark benchmark are used to provide some concept-of-proof analysis.
%\subsection{64 AMD CPU cores vs 1 NVIDIA A30}
%In this section, programs generated by \name{} are executed on AMD CPUs to compared with CUDA programs on NVIDIA A30 GPUs. 

\subsection{Roofline}
We use the Roofline model~\cite{williams2009roofline} to analyze the performance for both CPUs and GPUs. As CloverLeaf has complex interactions between the host (Fortran and C++) and the CUDA code, we use CUDA kernels from the Hetero-Mark. For the NVIDIA Ampere roofline, the effective FLOPs (green dots) are close to the theoretical memory-bandwidth upper bound (green curves), which indicates that the kernels have good memory access patterns. These kernels achieve the peak memory bandwidth and gain the upperbound FLOPs under these arithmetic intensities.
For both the x86 and Arm roofline, however, the dots are far below the curves, which indicates that there is a large gap between the effective memory bandwidths and the peak bandwidth. Specifically, for PR, FIR, EP, and KMeans examples, the CUDA real FLOPs are close to the upper bounds, while the transformed CPU programs are far below the bounds. This indicates that memory access patterns that are friendly for GPUs may be transformed to low-efficiency memory access patterns with poor locality for CPU architectures.

  \begin{figure}[htbp]
    \centering
    \includegraphics[width=80mm]{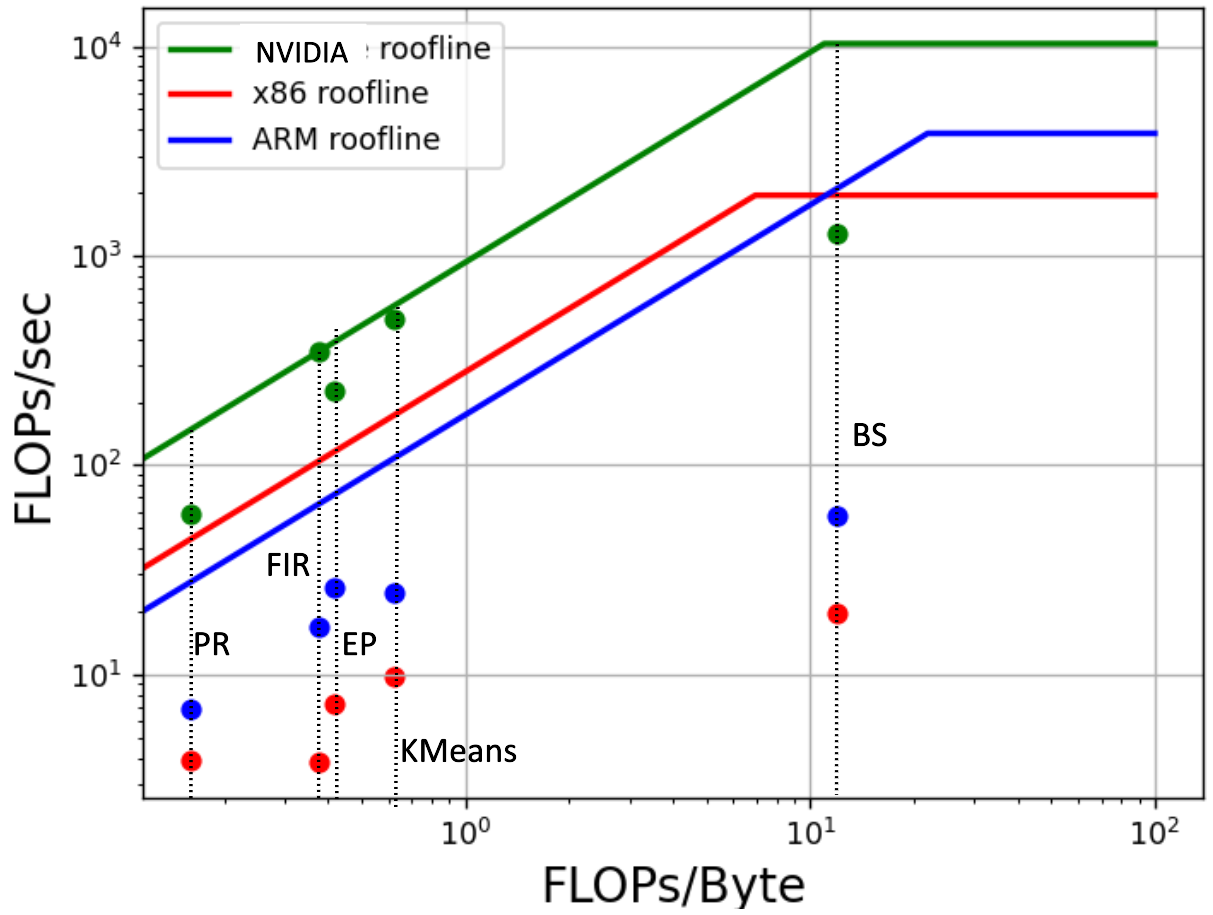}
    \caption{Roofline models for x86 CPU, AArch64 CPU (Server-AMD-A30, Server-Arm2), and NVIDIA Ampere GPU (Server-AMD-A30). The performance only includes kernel execution and does not include data movement between host and device.}
    \label{fig:roofline}
  \end{figure} 

\ignore{
The execution time is shown in Figure~\ref{fig:amd_cuda}. On average, \name{} is 16x slower than CUDA for kernel execution time.

  \begin{figure}[htbp]
    \centering
    \includegraphics[width=60mm]{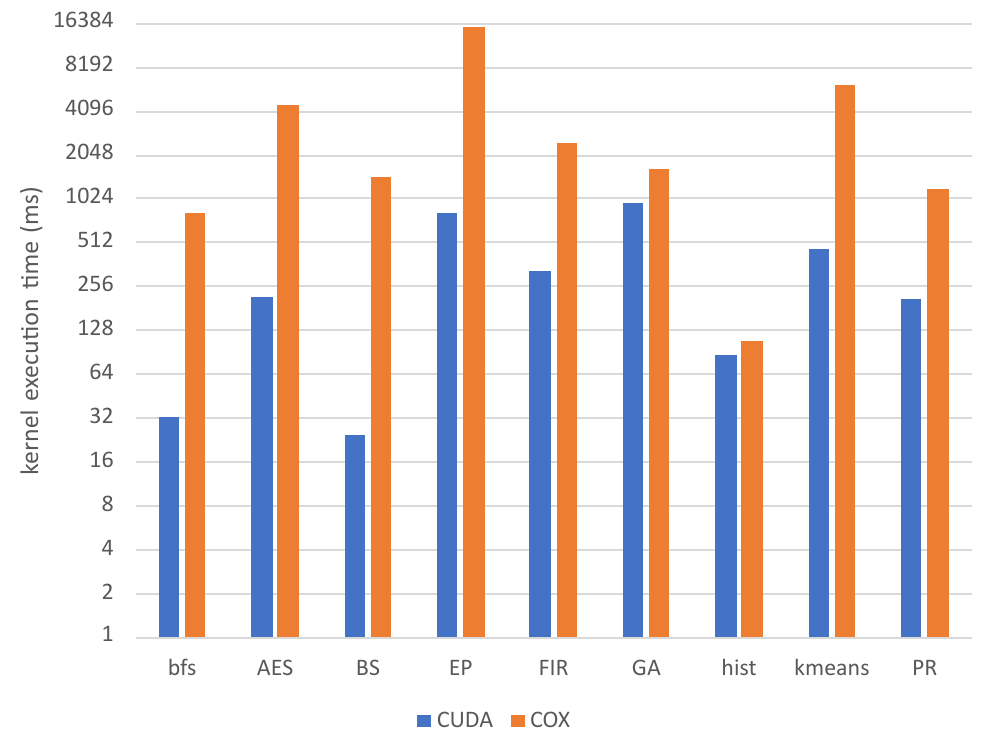}
    \caption{}
    \label{fig:amd_cuda}
  \end{figure} 

Same like ~\cite{svedin2021benchmarking}, we introduces Roofline model to analyze these benchmarks.
\subsection{Case study}
}

\subsection{Memory access reordering}
\begin{figure*}
  \centering
    \begin{subfigure}[t]{0.30\textwidth}
      \includegraphics[width=\textwidth]{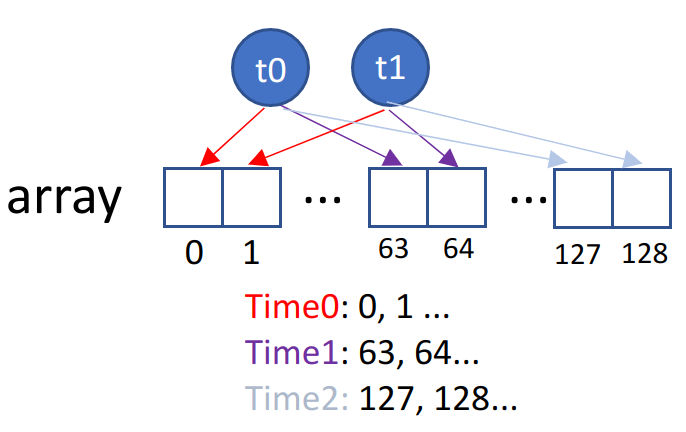}
      \caption{GPU memory access pattern.}
    \end{subfigure}
    %\\
    \begin{subfigure}[t]{0.30\textwidth}
      \includegraphics[width=\textwidth]{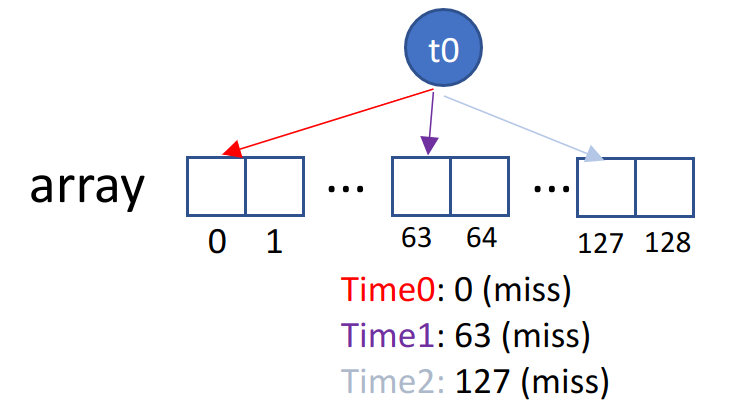}
      \caption{CPU memory access pattern after transformation. }
    \end{subfigure}
    %\\
    \begin{subfigure}[t]{0.30\textwidth}
      \includegraphics[width=\textwidth]{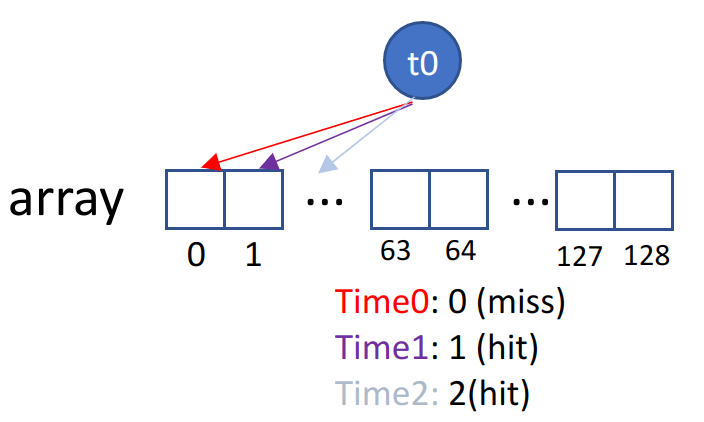}
      \caption{CPU memory access pattern after reordering.}
    \end{subfigure}
    \caption{Memory access patterns in GPU, CPU and after reordering in CPU.}
    \label{fig:hist_opt}
  \end{figure*}

To further understand the performance gap between GPUs and CPUs, we analyze the memory access patterns in the HIST program from Hetero-Mark. 
Figure~\ref{fig:hist_opt}(a) shows that 
%In this section, cases study are used to demonstrate the difference between GPU and CPU memory access, and potential solutions to speed up the CPU programs transformed by \name{}.  \\
%For the kernel function in Hetero-mark hist program, within a block, 
each GPU thread accesses memory address with a large stride. This pattern enables coalesced memory accesses within a GPU warp (i.e., all memory requests within a warp become sequential) 
 However, this feature will cause a low cache hit rate when the kernel is transformed to a CPU program (Figure~\ref{fig:hist_opt}(b)). By simply reordering the memory access sequence (Figure~\ref{fig:hist_opt}(c)), the CPU program can achieve a much higher hit rate. A similar transformation can also be used to speed up other examples like the GA. With manual memory access transformation, the LLC hit rate can be significantly improved, as shown in Table~\ref{table:reordering_prof}.

\begin{table}[]
\centering
\resizebox{\columnwidth}{!}{%
\begin{tabular}{|c|c|c|c|c|c|}
\hline
                      & reordering? & \begin{tabular}[c]{@{}c@{}}LLC-loads\\  (1e9)\end{tabular} & \begin{tabular}[c]{@{}c@{}}LLC-load \\ misses (1e9)\end{tabular} & \begin{tabular}[c]{@{}c@{}}LLC-stores \\ (1e9)\end{tabular} & \begin{tabular}[c]{@{}c@{}}LLC-store\\ misses (1e9)\end{tabular} \\ \hline
\multirow{2}{*}{HIST} & yes         & 359                                                        & 165                                                              & 152                                                         & 38                                                               \\ \cline{2-6} 
                      & no          & 37290                                                      & 26656                                                            & 2999                                                        & 62                                                               \\ \hline
\multirow{2}{*}{GA}   & yes         & 133                                                        & 13                                                               & 80                                                          & 11                                                               \\ \cline{2-6} 
                      & no          & 492                                                        & 148                                                              & 636                                                         & 446                                                              \\ \hline
\end{tabular}%
}
\caption{LLC access pattern differences between memory reordering or not in \name{}.}
\label{table:reordering_prof}
\end{table}

Besides cache locality, vectorization is another critical component, as we discussed for the 
EP, KMeans, and NW benchmarks (Section~\ref{sec:performance}). Without vectorization, CPU code cannot achieve peak performance, whereas a GPU can still take advantage of parallelization within a warp. Improving vectorization after code transformation is our future work. 

%can also improve performance. The implicitly lock step in CUDA warps can bring opportunity for SIMD vectorization. However, the programs transformed by \name{} are too complex to be vectorized by compilers. Future works are needed for auto vectorization~\cite{tian2017llvm,masten2018function,larsen2000exploiting,moll2018partial}.

\subsection{Runtime system optimization}
Other critical concerns for performance are kernel launches and synchronization. NVIDIA GPUs have a hardware scheduler to dispatch CUDA thread blocks to the streaming multiprocessors. Meanwhile, \name{} relies on the OS thread scheduler to manage the multi-core in CPUs and uses conditional variables to implement synchronization. As a result, \name{} has larger context switch workloads when there is a high number of kernel launches and a large amount of synchronization (Figure~\ref{fig:normalized_intel}).
Hence, to improve the runtime system performance, thread scheduling and synchronization has to be optimized with OS support in mind.

  \begin{figure}[htbp]
    \centering
    \includegraphics[width=80mm]{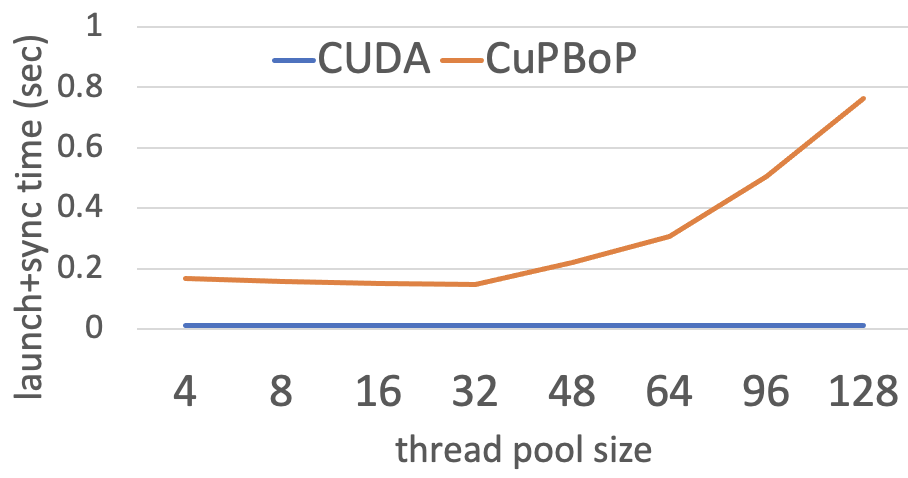}
    \caption{Execution time of 1000 kernel launches + synchronization on Server-Intel.}
    \label{fig:normalized_intel}
  \end{figure} 
  
\ignore{
To avoid the impact of runtime system, only CUDA programs with relatively fewer kernel launch (<1000) and large kernels are concerned for comparison. After manually optimization for the CPU programs, \name{} on x86 CPU is 12x slower than CUDA on GPU (Table~\ref{table:AMD-A30}), much smaller than the peak FLOPs difference (50x) between CPU and GPU.

\begin{table}[]
\centering
\begin{tabular}{|c|c|c|c|}
\hline
name   & GPU (sec)       & CPU (sec)    & CPU/GPU     \\ \hline
BS     & 1.09228   & 69.337  & 63.47914454 \\ \hline
EP     & 0.5754    & 8.64317 & 15.0211505  \\ \hline
kmeans & 1.4897    & 68.3919 & 45.90984762 \\ \hline
PR     & 1.86599   & 18.3448 & 9.831135215 \\ \hline
bfs    & 0.0679579 & 1.747   & 25.70709219 \\ \hline
GA     & 0.93      & 2.932   & 3.152688172 \\ \hline
hist   & 1.74134   & 2.04278 & 1.173108066 \\ \hline
Avg    &           &         & 12.22972072 \\ \hline
\end{tabular}
\caption{Kernel execution time (sec) on the x86 CPU and the NVIDIA Ampere GPU.}
\label{table:AMD-A30}
\end{table}
}

\section{Related work~\label{sec:related_work}}

\begin{table*}[]
\centering
\resizebox{180mm}{18mm}{%
\begin{tabular}{|c|l|cccc|}
\hline
Project(s) & Translate from                & \multicolumn{1}{c|}{Translate host programs?} & \multicolumn{1}{c|}{active} & \multicolumn{1}{c|}{Runtime implementation}      & output                                      \\ \hline
MCUDA~\cite{stratton2008mcuda}      & \multirow{6}{*}{CUDA source}  & \multicolumn{1}{c|}{no}                       & \multicolumn{1}{c|}{no}         & \multicolumn{1}{c|}{OpenMP}                      & C/C++                                       \\ \cline{1-1} \cline{3-6} 
Cumulus~\cite{blomkvist2021cumulus}    &                               & \multicolumn{1}{c|}{yes}                      & \multicolumn{1}{c|}{yes}        & \multicolumn{1}{c|}{no multi-thread}             & C/C++                                       \\ \cline{1-1} \cline{3-6} 
Swan~\cite{harvey2011swan}       &                               & \multicolumn{1}{c|}{no}                       & \multicolumn{1}{c|}{no}         & \multicolumn{1}{c|}{OpenCL}                      & OpenCL                                      \\ \cline{1-1} \cline{3-6} 
CU2CL~\cite{martinez2011cu2cl} and ~\cite{sathre2019portability,perkins2017cuda}      &                               & \multicolumn{1}{c|}{yes}                      & \multicolumn{1}{c|}{yes}        & \multicolumn{1}{c|}{OpenCL}                      & OpenCL                                      \\ \cline{1-1} \cline{3-6} 
HIPIFY~\cite{HIPIFY}     &                               & \multicolumn{1}{c|}{yes}                      & \multicolumn{1}{c|}{yes}        & \multicolumn{1}{c|}{HIP framework}               & HIP                                         \\ \cline{1-1} \cline{3-6} 
DPCT~\cite{DPCT}       &                               & \multicolumn{1}{c|}{yes}                      & \multicolumn{1}{c|}{yes}        & \multicolumn{1}{c|}{Intel OneAPI}                & DPC++                                       \\ \hline
Ocelot~\cite{diamos2010ocelot}     & \multirow{3}{*}{PTX Assembly} & \multicolumn{1}{c|}{yes}                      & \multicolumn{1}{c|}{no}         & \multicolumn{1}{c|}{Hydrazine threading library} & Binary code for AMD GPU, NVIDIA GPU and CPU \\ \cline{1-1} \cline{3-6} 
Horus~\cite{elhelw2020horus}      &                               & \multicolumn{4}{c|}{simulator}                                                                                                                                                   \\ \cline{1-1} \cline{3-6} 
GPGPU-Sim~\cite{gpgpu-sim}  &                               & \multicolumn{4}{c|}{simulator}                                                                                                                                                   \\ \hline
COX~\cite{han2021cox}        & \multicolumn{1}{c|}{NVVM IR}  & \multicolumn{1}{c|}{no}                       & \multicolumn{1}{c|}{yes}        & \multicolumn{1}{c|}{not provided}                    & Kernel code only                     \\ \hline
{\bf \name{}}        & \multicolumn{1}{c|}{NVVM IR}  & \multicolumn{1}{c|}{yes}                       & \multicolumn{1}{c|}{yes}        & \multicolumn{1}{c|}{portable run-time system}                    & Binary code for CPU                         \\ \hline
\end{tabular}%
}
\caption{Summary of executing CUDA to non-NVIDIA device.}
\label{table:related_work}
\end{table*}

Supporting multiple hardware device with one program is not a new topic. Many frameworks and programming languages have been proposed for portable programming between different architectures. However, most studies~\cite{jin2021evaluating,castano2021intel} focus on GPU-to-GPU migration, which does not involve SPMD-to-MPMD transformation and runtime library implementation. In this section, only migrations from GPUs to CPUs are of concern. Table~\ref{table:related_work} shows the summary of projects in this space. 
\subsection{SPMD to MPMD}
There is a huge gap on thread parallelism degree between CPUs and GPUs. To execute all these GPU threads by corresponding CPU threads, CPUs have to create more threads than their parallelism degree, which will create a lot of context switching and significantly decrease the performance. MCUDA~\cite{stratton2008mcuda} uses a for-loop (named thread loop) to serialize logical threads within a block (shown in Code~\ref{code:cpu_program}); this mechanism is also proposed in ~\cite{stratton2010efficient,stratton2013performance}. With this transformation, each CPU task has more workload, and the total number of CPU tasks decreases, so the context switching overload becomes negligible. However, when there are synchronization instructions within the kernel, naively wrapping all threads together will generate incorrect results. Several projects~\cite{shirako2009chunking,jaaskelainen2015pocl,zhang2013improving} discuss the issue of how to support the transformation when there are synchronization instructions within conditional sentences (e.g., for-loop, while-loop, if-else sentences). The previous CUDA version has an implicit warp-level barrier, as threads with a warp will always be executed in a lock step. The author in~\cite{guo2011correctly} highlights the importance of maintaining these lock-step features while migrating CUDA to CPU devices. All previous projects only use a single for-loop to serialize the threads, which is not enough to support warp-level functions (e.g., warp-shuffle, warp vote). Thus, researchers in \cite{han2021cox} use two-level nested for-loops to wrap the GPU kernels, where the outer for-loop stands for warps within a block and the inner for-loop stands for threads within a warp. However, COX does not have a high-performance runtime system, as it incurs thread create/join for each kernel launch. Besides, it does not support host code compilation, so it only produces kernel code and requires developers to manually rewrite the host code. 
%As all GPU threads share the same code, the iteration among the generated for-loop may also access the same/near memory, which bring spatial and temporal locality. 

The authors in \cite{zhang2013improving,karrenberg2012improving} proposed vectorizing the generated for-loop by using the static analysis. The C extensions for array notation are also used to accelerate the generated CPU programs~\cite{stratton2013performance}. 
Besides software transformations, some researchers tried to modify the CPU architecture in the hardware~\cite{chen2018enabling} or system level~\cite{gummaraju2010twin} to eliminate the time for context switching. These optimizations are beyond the scope of this paper.
In this paper, two runtime frameworks to run SPMD programs on CPU are used for evaluation with \name{}.
\subsubsection{DPC++ runtime}
DPC++ runtime is based on OpenCL, which allows different hardware support. POCL~\cite{jaaskelainen2015pocl} is the only open-source implementations that supports CPU backend. With the given OpenCL kernels, POCL applies the SPMD-to-MPMD transformtion to get MPMD kernels. The same as \name{}, it maintains a thread pool and task queue to support launching and executing these kernels. Besides, POCL supports JIT compilation. Instead of maintaining block size and grid size as variables in MPMD kernels, POCL replaces these variables with actual values during the kernel launch. Although this may lead to a higher latency of JIT compilation, it makes MPMD kernels easy to optimize.
\subsubsection{HIP-CPU runtime}
Different from \name{} and DPC++, HIP-CPU runtime does not involve compilations for SPMD-to-MPMD transformation. It provides libraries for both HIP kernel functions and HIP runtime functions. It maps GPU threads to CPU fiber instead of iterations in the generated for-loop. Thus, it has a higher overhead for context switching. Besides, as HIP-CPU cannot apply compilation-level transformations, its performance may be slow. For example, HIP-CPU needs to synchronize all device threads before memory movement between host and devices, regardless of whether or not these device threads will read/write this memory. However, HIP-CPU runtime is easy to maintain and is based on C++ languages. Thus, as long as a platform supports C++, it can execute HIP-CPU.

\subsection{CUDA to non-NVIDIA device}
%The CUDA compilation process~\cite{wu2016gpucc} is shown in Figure~\ref{fig:clang_compile}. 

To the best of our knowledge, most previous projects that migrate CUDA to non-NVIDIA devices can be grouped in two classes: source-to-source transformation and PTX assembly execution. \name{} is  the first project to support CUDA by LLVM IR and NVVM IR and provides a ready-to-run binary. Although Cuda-on-CL~\cite{perkins2017cuda} also compiled source code to LLVM IR (bytecode), it does not have SPMD-to-MPMD transformation. Instead, it directly converts the bytecode back to high-level source code (OpenCL). POCL~\cite{jaaskelainen2015pocl} implements SPMD-to-MPMD transformation on LLVM IR, but it only supports OpenCL programs. COX~\cite{han2021cox} applies SPMD-to-MPMD transformation on NVVM IR, but it does not compile the host program and thus no final binary to run.

\section{Future work}
\subsection{Coverage}
Some NVIDIA-GPU applications rely heavily on CUDA libraries (e.g., cuDNN, cuBLAS). Currently, these CUDA libraries are not supported in \name{}. To support these libraries, the functions from these libraries need to be replaced with non-NVIDIA device implementations that map to the target device. For example, we can use Intel MKL-DNN~\cite{IntelMKLDNN} to replace cuDNN when executing on Intel devices while using ARM Compute Library~\cite{armComputeLibrary} for ARM devices. A high-performance and scalable implementation for these CUDA libraries remains as future work.
\subsection{Performance}
To optimize the end-to-end execution time, two challenges need to be solved.
First is how to generate high-performance MPMD programs. From the discussion in Section~\ref{sec:future}, we know 
the memory access patterns translated from SPMD programs may have poor locality and prevent the programs from reaching the peak FLOPs. Although several compiler optimizations~\cite{yi2004transforming,wolf1991data,carr1994compiler,mckinley1996improving} for improving locality have been proposed, the effectiveness of these optimizations on transformed MPMD programs needs future evaluation. Besides, \name{} cannot fully utilize the SIMD instructions. As CUDA threads with a CUDA block are always independent of each other, the iterations in the generated for-loop are also independent. For example, in Section~\ref{sec:background}, the {\tt vecADD} function in Listing ~\ref{code:cpu_program} can be optimized by SIMD instructions. \name{} currently relies on LLVM's vectorization optimizations to utilize SIMD instructions. For complex kernel programs, the existing optimization path cannot fully exploit these optimization opportunities. Utilizing the existing auto-vectorization~\cite{nuzman2006multi,karrenberg2015whole,maleki2011evaluation,nuzman2006auto,nuzman2006autovectorization} research or proposing new algorithms to vectorize the transformed MPMD programs remains as future work.
Another challenge is how to speed up the runtime system. Integrating the existing high-performance concurrent computing implementations~\cite{moodycamel, intelTBB, cpp11Multicore} into \name{} may speed up the current \name{} runtime system. However, taking into account that CUDA uses hardware schedulers to control CUDA SMs, the gap between software schedulers and hardware schedulers is still a challenge and remains as further research.

\section{Conclusion}
Most existing works to execute CUDA on non-NVIDIA devices use a source-to-source translator to translate CUDA to portable programming languages. However, this work demonstrates some of limitations of these approaches through the design and evaluation of our \name{} framework. %Specifically, we note that CUDA kernels typically rely on lightweight threads that are hardware scheduled and use strided memory accesses that do not map well to traditional cache-oriented CPU architectures. Despite these limitations, we propose specific optimizations in \name{} to combine tasks in both an average and aggressive block fetching mechanism as well as possible memory reordering techniques to improve the performance of transformed CUDA codes on a variety of CPU platforms. 
This work demonstrates the clear benefits of combining specific NVVM and LLVM IR optimizations into a code transformation tool for better coverage and performance of executing CUDA codes on non-NVIDIA devices. Our evaluation also illustrates areas for improvement needed to achieve peak performance across CPU platforms. We plan to specifically focus on better mappings for memory locality, support for compiler-driven SIMD vectorization, continued addition of features to support unique CUDA instructions/libraries, and integration of existing concurrent computing projects into the runtime system. With these continued improvements, we believe that \name{} will be a compelling option for HPC programmers as well as a unique compiler optimization framework for investigating how to run HPC CUDA codes on non-NVIDIA devices.

\newpage
\newpage
\bibliographystyle{IEEEtranS}

\bibliography{ref}

\newpage
\begin{appendices}
\section{Hardware configuration}
\label{sec:appendix_hw}
\ignore{
Four different servers are used to evaluate. Server {\tt A} contains 16 Intel Rocket Lake i7-11700 CPU and one NVIDIA GeForce GTX 1660 Ti GPU with Ubuntu 20.04 system. And the memory size is 16GB. This server is used to evaluate the CUDA execution time as the baseline. Server {\tt B} contains 64 Intel Cascade Lake Gold 6226R CPU core on Ubuntu 18.04 with 400GB memory. This server is used to evaluate the performance of \name{}, DPC++ and HIP-CPU on Intel devices. Server {\tt C} contains 48 ARM A64FX CPU core on Apollo 80 system with 32GB memory, used for evaluating frameworks on ARM device. Server {\tt D} contains 4 SiFIve U74-MC cores, which is based on RISC-V ISA. And the memory size is 16GB.
}
The following information records the software version:
\begin{itemize}
    \item Benchmark: Rodinia Suite 3.1, Hetero-Mark, CUDA SDK 10.1. 
    \item Software: DPC++ 2021.3.0, HIP-CPU(commit:56f559c9)
    \item Dependent library: TBB (2020.1.2), GNU GCC/G++ 11.1.0, LLVM 10.0, CUDA 10.1.
\end{itemize}

\section{Evaluation configuration\label{sec:appendix_sw}}
Data scale recorded in Table~\ref{table:intel_data_scale} is used to execute evaluation for Figure~\ref{fig:normalized_intel}.
\begin{table}[]
\begin{tabular}{|c|c|c|}
\hline
benchmark                    & samples        & problem size                       \\ \hline
\multirow{14}{*}{Rodinia}    & b+tree         & 1M elements                        \\ \cline{2-3} 
                             & backprop       & 65536 input nodes                  \\ \cline{2-3} 
                             & bfs            & graph1MW\_6.txt                    \\ \cline{2-3} 
                             & gaussian       & matrix1024.txt                     \\ \cline{2-3} 
                             & hotspot        & 1024 * 1024                        \\ \cline{2-3} 
                             & hotspot3D      & 512 * 512                          \\ \cline{2-3} 
                             & lud            & 2048                               \\ \cline{2-3} 
                             & myocyte        & 100 time steps                     \\ \cline{2-3} 
                             & nn             & 1280k, 5 nearest neighbors         \\ \cline{2-3} 
                             & nw             & 8000 * 8000 data points            \\ \cline{2-3} 
                             & particlefilter & -x 128 -y 128 -z 10 -np 1000       \\ \cline{2-3} 
                             & pathfinder     & 100000 * 1000 * 20                 \\ \cline{2-3} 
                             & srad           & 8192* 8192                         \\ \cline{2-3} 
                             & streamcluster  & 65536 points 256 dimensions        \\ \hline
\multirow{8}{*}{Hetero-Mark} & AES            & 1GB                                \\ \cline{2-3} 
                             & BS             & num-elements 2097152               \\ \cline{2-3} 
                             & ep             & max-generation 100 population 1024 \\ \cline{2-3} 
                             & fir            & num-data-per-block: 4096           \\ \cline{2-3} 
                             & ga             & 65536\_1024.data                   \\ \cline{2-3} 
                             & hist           & num-pixels 4194304                 \\ \cline{2-3} 
                             & kmeans         & 100000\_34.txt                     \\ \cline{2-3} 
                             & PR             & 8192.data                          \\ \hline
\end{tabular}
\caption{Datascale for Figure~\ref{fig:normalized_intel}.}
\label{table:intel_data_scale}
\end{table}

Table~\ref{table:arm_riscv_data_scale} records the data scale used to get Figure~\ref{fig:Arm_riscv}.

\begin{table}[]
\begin{tabular}{|c|c|c|}
\hline
ISAs                     & samples & problem size                    \\ \hline
\multirow{8}{*}{AArch64} & AES     & 64MB                            \\ \cline{2-3} 
                         & BS      & num-elements 2097152            \\ \cline{2-3} 
                         & ep      & max-generation 5 population 256 \\ \cline{2-3} 
                         & fir     & num-data-per-block: 256         \\ \cline{2-3} 
                         & ga      & 65536\_1024.data                \\ \cline{2-3} 
                         & hist    & num-pixels 4194304              \\ \cline{2-3} 
                         & kmeans  & 100000\_34.txt                  \\ \cline{2-3} 
                         & PR      & 8192.data                       \\ \hline
\multirow{8}{*}{RISC-V}  & AES     & 64MB                            \\ \cline{2-3} 
                         & BS      & num-elements 2097152            \\ \cline{2-3} 
                         & ep      & max-generation 5 population 256 \\ \cline{2-3} 
                         & fir     & num-data-per-block: 256         \\ \cline{2-3} 
                         & ga      & 65536\_1024.data                \\ \cline{2-3} 
                         & hist    & num-pixels 4194304              \\ \cline{2-3} 
                         & kmeans  & 100000\_34.txt                  \\ \cline{2-3} 
                         & PR      & 8192.data                       \\ \hline
\end{tabular}
\caption{Datascale for Figure~\ref{fig:Arm_riscv}.}
\label{table:arm_riscv_data_scale}
\end{table}
\end{appendices}

\end{document}